\documentclass[journal]{IEEEtran}
\usepackage{amssymb,amsthm}       
\usepackage{multicol}
\usepackage[english]{babel} 

\usepackage{graphics} 
\usepackage{epsfig} 
\usepackage{times} 
\usepackage{amsmath} 
\usepackage{algorithm} 
\usepackage{color}

\usepackage{cite}
\usepackage{caption}
\usepackage{subcaption}
\usepackage{setspace}
\usepackage{bm} 
\usepackage{pifont}
\newcommand{\argmax}{\arg\!\max} 
\newcommand{\argmin}{\arg\!\min}
\newtheorem{result}{\bf Result}

\newcommand*{\rom}[1]{\expandafter\@slowromancap\romannumeral #1@}

\title{Noisy One-bit Compressed Sensing with Side-Information 
}

\author{ 
	\IEEEauthorblockN{Swatantra Kafle, \textit{Student Member, IEEE}, Thakshila Wimalajeewa, \textit{Senior Member, IEEE,} and Pramod K. Varshney, \textit{Life Fellow, IEEE}}
}

\begin{document}

	\maketitle
	\thispagestyle{empty}
	\pagestyle{empty}

	\begin{abstract}
		We consider the problem of sparse signal reconstruction from noisy one-bit compressed measurements when the receiver has access to side-information (SI). We assume that compressed measurements are corrupted by additive white Gaussian noise before quantization and sign-flip error after quantization.  A generalized approximate message passing-based method for signal reconstruction from noisy one-bit compressed measurements is proposed which is then extended for the case where the receiver has access to a signal that aids signal reconstruction, i.e., side-information. Two different scenarios of side-information are considered-a) side-information consisting of support information only, and b) side information consisting of support and amplitude information. SI is either a noisy version of the signal or a noisy estimate of the support of the signal.  We develop reconstruction algorithms from  one-bit measurements using noisy SI available at the receiver. Laplacian distribution and Bernoulli distribution are used to model the noise which when applied to the signal yields the SI for the above two cases. The Expectation-Maximization algorithm is used to estimate the noise parameter using noisy one-bit compressed measurements and the SI. We show that one-bit compressed measurement-based signal reconstruction is quite sensitive to noise, and the reconstruction performance can be significantly improved by exploiting available side-information at the receiver.
	\end{abstract}
	
	\begin{keywords}
		sparse signal reconstruction, one-bit compressed measurement,  Generalized Approximate Message Passing, side-information   
	\end{keywords}
	\section{ Introduction}
	
	With the introduction of compressed sensing (CS) \cite{candes, donoho}, several algorithms have been proposed either for signal reconstruction from its low-dimensional measurements \cite{OMP, BCS, CSBP, Cevher1, IHT} or for inference tasks such as detection, estimation, and classification with or without reconstructing the signals \cite{Nowak, kafleDet, wimalajeewapartial, app}. 
	All these works assume that the measurements are real-valued. However, in most practical applications, quantization of the compressed measurements is required before transmission and/or storage. Some works \cite{quantized_dai, quantized_jacques} have addressed the quantized version of compressed sensing. Though real-valued compressed measurements can be approximated with high-rate quantization, coarse quantization is more attractive in practice as it significantly reduces bandwidth usage and power consumption. Among all the possible quantization schemes, one-bit quantization is highly preferred one where the measurements are quantized to their sign values. 
	The popularity of one-bit quantization is due to its simplicity, low-cost design, robustness to linear/non-linear distortion, and high sampling rate.
	
	One-bit CS\cite{boufounos, binarystableEmbedd, plan} deals with the reconstruction of sparse signals from one-bit quantized compressed measurements. It is attractive in sensor networks since it provides savings of scarce network resources such as communication bandwidth, transmit/processing power and storage.  It has been shown that in some practical scenarios, one-bit compressed sensing can outperform multi-bit quantized compressed sensing \cite{regimeCh_Laska}. Several reconstruction algorithms have been proposed that allow reconstruction from one-bit quantized measurements \cite{kafle2016, plan, musa, kamilov, binarystableEmbedd, baraniukexponential, normEstimation,classification_binaryGAMP, kafle2019noisy}. The works in\cite{kamilov, musa, plan} consider signal reconstruction from one-bit compressed measurements using a model which either does not consider any noise or considers only additive white Gaussian noise. Though one-bit compressed sensing has shown promise of decent inference and signal reconstruction performance, it has been shown to be quite sensitive to noise. Some works have addressed this issue by the use of multiple measurement vectors \cite{kafle}. We emphasize that these works consider the presence of Gaussian noise only.
	
	In this work, we consider a generalized measurement model of one-bit CS where noise is assumed to be added at two stages of the measurement process- a) before quantization and b) after quantization. We model the noise before quantization as additive white Gaussian noise and the noise after quantization as a sign-flip noise generated from a Bernoulli distribution. We approach the problem from the Bayesian perspective. Hence, we impose Bernoulli-Gaussian density as a prior on the signal to model sparse structure. Several works \cite{BCS, amp, CSBP } have addressed Bayesian compressed sensing with real-valued measurements, i.e., when the observations are linear and are corrupted by AWGN. Note that one-bit compressed measurements are highly non-linear, and noise in the measurement process is not AWGN. Hence, we consider using the generalized approximate message passing (\texttt{GAMP}) algorithm \cite{rangan}, an extension of the approximate message passing (\texttt{AMP}) algorithm \cite{amp}, as it provides a systematic approach to impose any prior on a signal and non-linearity in the measurements.  \texttt{GAMP} and \texttt{AMP} algorithms are popular as they provide an efficient iterative procedure to approximate the MMSE estimator, which are otherwise analytically intractable and computationally inefficient. 
	Further, these algorithms also allow the estimation of signal and channel parameters using  the Expectation-Maximization (EM) algorithm \cite{parameterEst_kamilov,parameterEst_vila} during signal reconstruction. Hence, we develop a noisy one-bit CS algorithm using the \texttt{GAMP} framework and we refer to the algorithm as \texttt{noisy1bG}.

	Next, we consider the problem where a receiver has access to a signal which is similar to the signal that we want to reconstruct from its noisy one-bit measurements. We refer to this signal as side-information (SI). SI is available in many applications, including in the reconstruction of sequences of signals such as in dynamic MRI reconstruction \cite{lu}, video signal reconstruction \cite{video}, and sequential estimation \cite{dyn}. 
	In this work, we aim to improve signal reconstruction performance of the one-bit CS algorithms by exploiting SI at the receiver.  
	
	Several works in the literature  \cite{side_mota,side_vaswani,side_wang, side_heterobay,side_chen,side_maAMP} have looked into ways of using/exploiting SI  to improve signal reconstruction performance. In \cite{side_vaswani}, the authors assume that the receiver has the knowledge of the partial support set of the sparse signal, whereas in \cite{side_mota, side_wang} it is assumed that SI is present at the receiver which is assumed to be a noisy version of the actual compressed signal. All of these works assume that the compressed measurements have infinite precision. The authors in \cite{side_biht} use one-bit compressed measurements and assumes that the receiver has access to the partial support set of the signal.  
	
	In this work, we develop algorithms based on the \texttt{GAMP} framework that exploit the two different kinds of SI available at the receiver to improve reconstruction performance. First, we assume that the receiver has access to side-information consisting of both support and amplitude information. This is usually the case when the receiver has access to the signal estimated or reconstructed at the previous time instant. Based on the temporal dynamics of the observed phenomenon, the support, and the amplitude of the sparse signal might change over time. Further, due to noise in the compressed measurement process, the reconstructed signals might have some incorrect support and amplitude information. Hence, we model the SI as the signal corrupted with additive noise to account for the discrepancies between the SI and the signal. For this setup, we develop two algorithms for two cases when we model the additive noise in SI using Laplacian distribution (referred to as \texttt{LaplacianSI}) and Gaussian distribution (referred to as \texttt{GaussianSI}) respectively, and study the reconstruction performance. We show that modeling noise using a sparsity promoting density has better reconstruction performance, i.e., \texttt{LaplacianSI} performs better than \texttt{GaussianSI}. Second, we assume that the receiver has access to the support information as SI. Recent work \cite{qing2019superimposed} considers the support information as the side-information for the one-bit compressed sensing problem. However, the authors assume that either the complete or partial support set of the SI lies in the support set of the compressed signal, which is usually not the case. In this work, we consider a general model in which we assume that the support set available at the receiver is a noisy version of the support set of the signal.  We model the error in the support information between the compressed signal and the SI by a Bernoulli distributed random variable. We, then, develop a \texttt{GAMP}-based algorithm (referred to as \texttt{SupportSI}) to reconstruct the sparse signal when the receiver has a noisy support set as the SI. In both cases, we use the Expectation-Maximization (EM) algorithm to estimate the noise parameters.


	
	

	We emphasize that the reconstruction performance of one-bit CS is susceptible to noise, and mitigating noise leads to better reconstruction performance. We show the improvement in performance through numerical simulations. Further, we show that incorporating side-information at the receiver leads to improved reconstruction performance. 
	
	The main contributions of the paper are 
	\begin{itemize}
		\item In contrast to most existing works where only either pre- or post-quantization noise is considered for the one-bit compressed sensing problem, we consider a model that includes both pre- and post-quantization noise and develop reconstruction algorithms using the GAMP framework. 
		\item We extend the proposed algorithm to the case where the receiver has access to SI. We consider two  possible scenarios for SI: a) SI with both support and amplitude information, and b) SI with only support information
		
		\item We provide closed-form expressions for the evaluation of all the non-linear equations in the GAMP algorithms for all the proposed algorithms. This makes the algorithms more time-efficient. 
		
		\item Through detailed numerical simulations, we show that the proposed methods yield improved reconstruction performance compared to the state-of-the-art algorithms. 
	\end{itemize}
	A part of this work was presented at Asilomar 2018 \cite{side_kafle}.
	\vspace{- 0.25 cm}
	\subsection*{Organization} 
	The paper is organized as follows. In Section II, the signal and the measurement models are defined for noisy one-bit CS and noisy one-bit CS with SI. GAMP based one-bit CS algorithm with pre-quantization, and post-quantization noise is developed in Section III. In Section IV, we develop algorithms for one-bit CS with SI having both support and amplitude information. We develop two different algorithms based on how the dependence of SI on the signal is modeled. In Section V, we develop an algorithm for the case when the receiver has a noisy support set as the SI. In Section VI, numerical results are presented to illustrate the performance of different proposed algorithms. Section VII concludes the paper.
	\subsection*{Notation}
	In this paper, we use the following notations. Scalars are denoted by lower case letters and symbols, e.g., $y$ and $\gamma$. Vectors and matrices are represented by lowercase boldface and uppercase boldface characters such as $\mathbf{x}$ and $\mathbf{A},$ respectively. Hadamard product, i.e., element wise product is denoted by $\odot$. We represent the Gaussian pdf with mean $m$ and variance $v$ by $\mathcal{N}(.;m,v)$. Similarly, we represent Laplacian pdf with mean and variance $v$ by $\mathcal{L}(.;m,v)$ We define $I_n(a,b; m, v) = \int_a^b x^n \mathcal{N}(x;m,v) dx$ and $PI_n(\tau,m,v) = \int x^n \Phi(x/\sqrt{\tau}) \mathcal{N}(x;m,v) dx.$ Further, $ \Phi(x) = I_0(x; 0, 1),$ and  $\phi(x) = \mathcal{N}(x;0,1)$.

	\section{Signal and Measurement Models}
	In the following, we introduce our signal and measurement models.
	\subsection{Signal Model}
	We consider the input signal $\mathbf{x} \in \mathbb{R}^N$ to be random with elements having identical and independent (i.i.d.) distribution 
	\begin{equation}
	p_{\bm{\mathcal{X}}}(\mathbf{x}) = \prod_{n=1}^{N} p_{\mathcal{X}_n}(x_n),
	\end{equation}
	where each component ${x_i}$ is a Bernoulli-Gaussian distributed random variable with pdf
	\begin{equation} \label{signalModel}
	p_{\mathcal{X}_n}(x_n) = (1-\lambda) \delta(x_n) + \lambda \mathcal{N}(x_n; 0, v_x),
	\end{equation}
	where $\delta(x)$ is  the Dirac-delta function, and  $\lambda$ is the probability of having non-zero values. $\mathbf{x}$ is a sparse signal. $\lambda$ controls the sparsity of the signal. Smaller the value of $\lambda$, sparser the signal.  
	
	\subsection{Measurement Model}
	\begin{figure}
		\centering
		\includegraphics[width=0.35\textwidth]{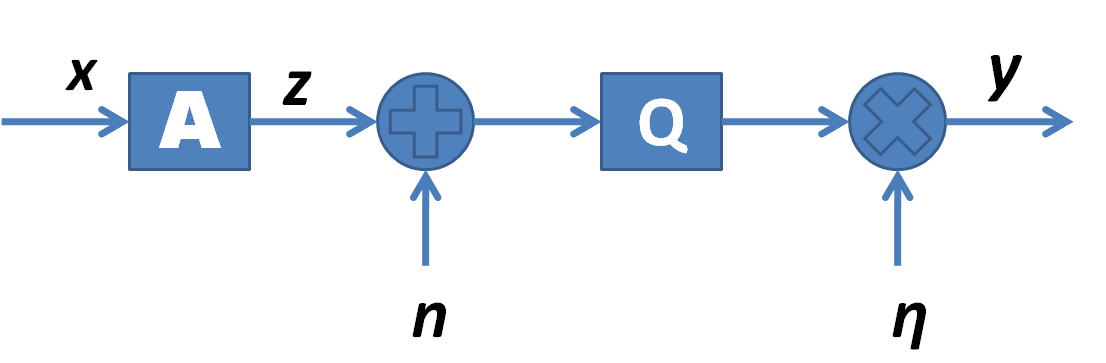}
		\caption{One-bit CS with pre-quantization and post-quantization noise.} \label{block}
	\end{figure}
	Figure \ref{block} shows the transmission chain of the measurement model of the problem considered in this work. The sparse signal $\mathbf{x} \in \mathbb{R}^N$ is linearly transformed to a vector $\mathbf z \in \mathbb{R}^M$ using the random measurement matrix $\mathbf A \in \mathbb{R}^{M \times N}$. The transformed vector, $\mathbf{z}$, is assumed to be corrupted by additive i.i.d Gaussian noise vector with mean zero and variance $v, \text{i.e.}, n_m \sim \mathcal{N}(0,v).$ This corrupted compressed vector is quantized element-wise to $+1$ or $-1$ based on the sign of the signal. We assume a noisy channel between the quantizer and the receiver where the quantized measurements are corrupted by multiplicative noise which takes values either $+1$ or $-1$.
	Formally, the quantized measurement model can be written as
	\begin{equation} \label{problemformulation}
	\mathbf{y} = \bm{\eta} \odot {Q}(\mathbf{Ax}+\mathbf{n}),
	\end{equation}
	where the quantizer ${Q} : \mathbb{R}^M \rightarrow \{-1,+1\}^M$ is the element-wise sign quantizer.
	The $m$-th element at the output of the quantizer is
	\begin{eqnarray}
	Q(\zeta_m) = \left\{
	\begin{array}{ccc}
	+1, ~&~ \textrm{if}~ \zeta_{m} > 0,\\
	-1, ~&\textrm{if}~ \zeta_{m} \leq 0,\\
	\end{array}\right.\label{quantization}
	\end{eqnarray}
	and $\bm{\eta} \in \{ -1, +1\}^M$ is the i.i.d. post-quantization noise. $\eta_m$ is assumed to follow Bernoulli distribution with $Pr(\eta_m = 1) = \gamma$. 
	We define the inverse of quantization function, $Q^{-1}(.),$ as
	\begin{eqnarray}
	Q^{-1}(y_m) = \left\{
	\begin{array}{ccc}
	(-\infty, 0 ~\rbrack, ~&~ \textrm{if}~ y_{m} \leq 0, \\
	(0, \infty), ~&\textrm{if}~ y_{m} > 0,\\
	\end{array}\right.\label{invquant1}
	\end{eqnarray}
	where $y_m$ is the $m$-th element of $\mathbf{y}.$
	\subsection{Noisy one-bit CS}
	As in \cite{kamilov}, the posterior distribution of the signal, $\mathbf{x}$, given the quantized and noisy measurements, $\mathbf{y}$, at the receiver is
	\begin{align}
	p_{\bm{\mathcal{X}}|\bm{\mathcal{Y}}}({\mathbf{x}|\mathbf{y}}) 
	&\propto p_{\bm{\mathcal{Y}}|\bm{\mathcal{X}}}(\mathbf{y}|\mathbf{x}) p_{\bm{\mathcal{X}}}(\mathbf x) \\
	&\propto \prod_{m=1}^{M} \mathbb{I}_{z_m \in \{Q^{-1}(\eta_m y_m) \} }  \prod_{n=1}^{N} p_{\mathcal{X}_n}(x_n),  \label{noisy}
	\end{align}
	where $\mathbb{I}_{(.)}$ represents the indicator function, and $\propto$ represents equality upto a proportional constant. 
	The minimum mean square error (MMSE) estimator of $\mathbf{x}$ is the mean of the posterior distribution, i.e.,  $\mathbb{E}_{\bm{\mathcal{X}}|\bm{\mathcal{Y}}}[\mathbf x| \mathbf y]$. Next, we assume that the receiver has access to side-information which is related to the signal of interest. The side-information is imposed as probability distribution. Let $\widetilde{\mathbf{x}}$ represent the side-information of signal $\mathbf{x}.$
	Here, we construct the posterior distribution of signal, $\mathbf{x}$, given the noisy one-bit compressed measurements $\mathbf{y}$ and side-information $\widetilde{\mathbf{x}}$ as 
	\begin{align}
	\begin{split}
	p_{\bm{\mathcal{X}}|\bm{\mathcal{Y}}, \bm{\widetilde{\mathcal{X}}}}(\mathbf{x}|\mathbf{y}, \widetilde{\mathbf{x}}) 
	\propto p_{\bm{\mathcal{Y}}|\bm{\mathcal{X}}}(\mathbf{y}|\mathbf{x})  p_{\bm{\widetilde{\mathcal{X}}}|\bm{\mathcal{X}}}(\widetilde{\mathbf{x}}|\mathbf{x}) \\
	\propto \prod_{m=1}^{M} \mathbb{I}_{z_m \in \{Q^{-1}(\eta_m y_m) \} }  \prod_{n=1}^{N} p_{\mathcal{X}_n|\widetilde{\mathcal{X}}_n}(x_n|\widetilde{x}_n),
	\end{split} \label{sideInf} 
	\end{align}
	where $p_{\mathcal{X}_n|\widetilde{\mathcal{X}}_n}(x_n | \widetilde{x}_n)$ is the conditional density function that gives the statistical characterization of the sparse signal when the side-information is given. The MMSE estimator of $\mathbf{x}$ with SI at the receiver is $\mathbb{E}_{\bm{\mathcal{X}}|\bm{\mathcal{Y}}, \bm{\mathcal{\widetilde{X}}}}[\mathbf x | \mathbf y, \mathbf{\widetilde{x}}]$. We note that the derivation of the MMSE estimators (\ref{noisy}) and (\ref{sideInf}) is intractable in direct form. Thus, we develop \texttt{GAMP}-based algorithms to approximate the MMSE estimator. 

	\section{Noisy one-bit CS Algorithm}
	In this section we begin with an introduction to the \texttt{GAMP} algorithm. \texttt{GAMP} algorithm \cite{rangan} is a generalization of the \texttt{AMP} algorithm \cite{amp}. Both \texttt{AMP} and \texttt{GAMP} algorithms apply loopy belief propagation in the bipartite graph under the Gaussian approximation for the involved messages. These methods fall under the Bayesian framework which assume  a prior distribution, $p_{\bm{\mathcal{X}}}(\mathbf x)$, on $\mathbf{x}$. The key idea in the Bayesian approach is to find the marginal posterior distributions $p_{\mathcal{X}_n|\bm{\mathcal{Y}}}(x_n|\mathbf{y})$ which could be used in minimum mean square error (MMSE) or maximum a posteriori (MAP) estimation of each $x_n$ as:
	\begin{eqnarray*}
		\widehat{x}_{n}^{\textsc{MAP}} &=& \argmax_{x_n}p_{\mathcal{X}_n|\bm{\mathcal{Y}}}(x_n|\mathbf{y}),\\
		\widehat{x}_{n}^{\textsc{MMSE}}&=&\argmin_{\widehat{x}_n}\mathbb{E}_{\mathcal{X}_n,\bm{\mathcal{Y}}}\big\{(x_n-\widehat{x}_n)^2\big\}\\
		&=&\mathbb{E}_{\mathcal{X}_n|\bm{\mathcal{Y}}}\{x_n|\mathbf{y}\}.
	\end{eqnarray*}
	\texttt{AMP} inherently assumes the prior of a signal to be Gaussian whereas \texttt{GAMP} offers the systematic approach of taking any prior of the signal into account during the denoising step. However, the evaluation of the true marginal distributions, $p_{\mathcal{X}_n|\bm{\mathcal{Y}}}(x_n|\mathbf{y})$, of a high-dimensional vector, $\mathbf x$, is analytically intractable and computationally prohibitive.  The \texttt{GAMP} algorithm implements loopy belief propagation and uses the central limit theorem with quadratic approximations to  approximate $p_{\mathcal{X}_n|\bm{\mathcal{Y}}}(x_n|\mathbf{y})$ to improve computational performance. 
	The \texttt{GAMP} algorithm uses the sum-product and max-sum belief propagation algorithms to compute MMSE and MAP estimators respectively. In the next section, we focus on the MMSE estimation problem corresponding to the posterior densities (\ref{noisy}) and (\ref{sideInf}). For detailed expositions on \texttt{AMP}, and \texttt{GAMP}, we refer the readers to \cite{amp} and \cite{rangan}. In this work, we  consider the sum-product version of the \texttt{GAMP} algorithm where we find the MMSE estimator of $\mathbf{x}$ corresponding to the posterior densities (\ref{noisy}) and (\ref{sideInf}).   

	\subsection{Noisy one-bit CS  ( \texttt{Noisy1bG} )} 
	In this subsection, we develop a \texttt{GAMP} based algorithm that reconstructs a sparse signal from its noisy one-bit compressed measurements. Define $\mathbf z \triangleq \mathbf A \mathbf x$ as the linear transformation of $\mathbf x$. The transformed signal, $\mathbf{z}$, is corrupted by i.i.d. Gaussian noise which is quantized to one-bit as defined in  (\ref{quantization}). The one-bit quantized signal is transmitted over a channel with probability of sign-flip $1- \gamma$. We represent the entire effect of additive white Gaussian noise (measurement noise), one-bit quantization and sign-flip error (channel noise) by a probabilistic channel, $ p_{\bm{\mathcal{Y}}|\bm{\mathcal{Z}}}\left(\mathbf{y}|\mathbf{z};\sigma_w^2\right)$. Since we assume that the measurement noise and the channel noise are i.i.d., the channel is represented as 
	\begin{eqnarray}\label{channel}
	p_{\bm{\mathcal{Y}}|\bm{\mathcal{Z}}}\left(\mathbf{y}|\mathbf{z};\sigma_w^2 , \gamma\right)&\!\!=\!\!&\prod_{m=1}^Mp_{\mathcal{Y}_m|\mathcal{Z}_m}\left(y_m|z_m;\sigma_w^2, \gamma\right). 
	\end{eqnarray}
	
	In Algorithm 1, we summarize the steps of the \texttt{GAMP} algorithm for sparse signal reconstruction from one-bit noisy compressed measurements. We refer to this algorithm as \texttt{Noisy1bG}. This Algorithm requires the computations of non-linear functions $F_1(\cdot), F_2(\cdot), G_1(\cdot), \text{ and } G_2(\cdot)$ as defined in (\ref{measurementUp}) and (\ref{EstUp}). 
	
	\subsubsection*{Evaluation of $F_1(.)$ and $ F_2(.)$}
	First, we evaluate the channel, $p_{\mathcal{Y}_m|\mathcal{Z}_m} (y_m|z_m)$,  based on our system model as
	\begin{algorithm}
		\begin{enumerate}
			\item Initialization: Set t=0 and initialize $\widehat{\mathbf{x}}^t , \mathbf{\tau}_x^t,$ and $\widehat{ \mathbf{s}}^t$ as $\widehat{\mathbf{x}}^t = \mathbb{E}[\mathbf{x}],~~  \mathbf{\tau}_x^t = \text{var}[\mathbf{x}], ~~\widehat{ \mathbf{s}}^t = {0}, 
			$ where the expectation and variance of $\mathbf{x}$ are with respect to $p_x$
			\item \textbf{Measurement Update}
			\begin{itemize}
				\item Linear Step
				{ \small 
					\begin{align*}
					&\boldsymbol\tau^{t+1}_p = (\mathbf A \odot \mathbf A) \boldsymbol{\tau}_x^t,~~ 
					\widehat{ \mathbf{p}}^{t+1} =   \mathbf{A}\widehat{\mathbf{x}}^t -  \boldsymbol\tau^{p,t+1}\odot \widehat{ \mathbf{s}}^t,		 	     \end{align*} }
				\item Non-Linear Step
				{ \small \begin{align} \label{nonlinear1}
					\begin{split} 
					&\widehat{ \mathbf{s}}^{t+1}  =  F_1( \mathbf y, \widehat{ \mathbf{p}}^{t+1}, \boldsymbol\tau^{p,t+1}),\\ 
					&{ \boldsymbol{\tau}}^{t+1}_s =    F_2( \mathbf y, \widehat{ \mathbf{p}}^{t+1}, \boldsymbol\tau^{p,t+1}),
					\end{split}	 	   
					\end{align}}
				where $F_1$ and $F_2$ are applied element-wise and are defined as 
				{ \small  \begin{align} \label{measurementUp}
					\begin{split}
					& F_1 \big( y, \widehat{p}^{t+1}, \tau^{p,t+1}\big) = \frac{1}{\tau^{p,t+1}}\Big( \mathbb{E}[z|y] - \widehat{p}^{t+1} \Big),\\ 
					& F_2 \big(  y, \widehat{p}^{t+1}, \tau^{p,t+1} \big) = \frac{1}{\tau^{p,t+1}} \Big( 1 - \frac{\text{var}[z|y]}{\tau^p} \Big).
					\end{split}
					\end{align} }
				The expectation and variance are evaluated with respect to $z \sim \mathcal{N}(\widehat{p}, \tau^p)$.
			\end{itemize}
			\item \textbf{Estimation Update} 
			\begin{itemize}
				\item[] Linear Step
				{ \small 	\begin{align*}
					\boldsymbol\tau^{r,t+1} = 			((\mathbf A \odot \mathbf A)^T \boldsymbol{\tau}_s^{t+1})^{-1},~~ 
					\widehat{ \mathbf{r}}^{t+1} =  \widehat{ \mathbf{x}}^{t} + \boldsymbol{\tau}^{r,t+1} \odot(\mathbf A^T  \widehat{ \mathbf{s}}^{t+1}),		 	     \end{align*} }	
				where the inversion is performed element-wise		
				\item[] Non-linear Step
				{ \small 	\begin{align} \label{nonlinear2}
					\begin{split}
					&\widehat{ \mathbf{x}}^{t+1}  =  G_1( \widehat{ \mathbf{r}}^{t+1}, \boldsymbol\tau^{r,t+1};p_{\mathcal{X}}),\\ 
					&{ \boldsymbol{\tau}}^{t+1}_x =    G_2( \widehat{ \mathbf{r}}^{t+1}, \boldsymbol\tau^{r,t+1};p_{\mathcal{X}}),
					\end{split}	 	   
					\end{align} }
				where $ G_1$ and $ G_2$ are applied element-wise and are defined as 
				{ \small \begin{align} \label{EstUp}
					\begin{split}
					&G_1( \widehat{r}_n, \tau^r_n;p_{x_n})= \mathbb{E}_{\mathcal{X}_n|\bm{\mathcal{Y}}}[x_n| \mathbf y;\widehat{r}_n, \tau^r_{n}],
					\\&
					G_2( \widehat{r}_n,\tau^r_n;p_{x_n}) = \text{var}_{\mathcal{X}_n|\bm{\mathcal{Y}}}[x_n| \mathbf y;\widehat{r}_n,  \tau^r_{n}].
					\end{split}
					\end{align} }
			\end{itemize}
			The expectation and variance are evaluated with respect to $p_{\mathcal{X}_n|\bm{\mathcal{Y}}} \propto \mathcal{N}(\cdot; \widehat{r}_n, \tau^r_n) p_{\mathcal{X}_n}(\cdot).$ 
			\item []  Set t = t+1 and return to step  2 until $t<T$.	
			\caption{\texttt{Noisy1bG} Algorithm}\label{dalgo1}
		\end{enumerate} 
	\end{algorithm}

	\begin{align}
	\begin{split}
	& p_{\mathcal{Y}_m|\mathcal{Z}_m}( {y_m|z_m})  =  
	\sum_{y_{qm}} p(y_{qm}|z_m) p(y_m|y_{qm},z_m)  
	\\       & =   \gamma p(y_{qm}: y_{mq} = y_m  |z_m) + (1 - \gamma) p(y_{qm}: y_{qm} \neq y_m  |z_m),
	\end{split}
	\end{align}
	where $y_{qm}$ is the $m$-th element of the output of the quantizer $ \mathbf y_q$. Let $\delta_m^+ = \delta(y_m+1)$, and $\delta_m^- = \delta(y_m-1)$ . 
	It is noted that $ p(y_{qm}: y_{qm} = y_m  |z_m)$ is given by
	
	\begin{align} \label{ygivenz1}
	\begin{split}
	&p(y_{qm}: y_{qm} = y_m  |z_m) 
	\\ & =  p(z_m+n_m \geq 0 |z_m) \delta_m^- + p(z_m+n_m \leq 0 |z_m) \delta_m^+  
	\\ & =  \Phi(\frac{z_m}{\sqrt{v}}) \delta_m^- + ( 1 - \Phi(\frac{z_m}{\sqrt{v}})) \delta_m^+.
	\end{split}
	\end{align}
	\noindent Similarly, 	we evaluate $ p(y_{qm} : y_{qm} \neq y_m  |z_m)$ as
	\begin{align}\label{ygivenz2}
	\begin{split}
	&p(y_{qm} : y_{qm} \neq y_m  |z_m) 
	= (1 - \Phi(\frac{z_m}{\sqrt{v}}) )\delta_m^- + \Phi(\frac{z_m}{\sqrt{v}}) \delta_m^+.
	\end{split}
	\end{align} 
	Using central limit theorem arguments, \texttt{GAMP} approximates the distribution of random variable $\mathcal Z$ as Gaussian with mean $\widehat{p}$ and variance $\tau^p$, i.e., $\mathcal Z \sim \mathcal{N} (\widehat{p},\tau^p)$. The posterior marginal distribution, $p_{\mathcal{Y}_m|\mathcal{Z}_m} (y_m|z_m)$, can be evaluated as 
	\begin{eqnarray*}
		p_{\mathcal{Z}_m|\bm{\mathcal{Y}}}\big(z_m|\mathbf{y};\widehat{p}_m,\tau^p_{m}\big)=\frac{p_{\mathcal{Y}_m|\mathcal{Z}_m}(y_m|z_m)\mathcal{N}(z_m;\widehat{p}_m,\tau^p_{m})}{\int_{z_m}p_{\bm{\mathcal{Y}}|\mathcal{Z}_m}(y_m|z_m)\mathcal{N}(z_m;\widehat{p}_m,\tau^p_{m})}.
	\end{eqnarray*}
	The term in the denominator is the normalization constant. In the following, we evaluate the normalization constant, $Z^p_m$, the posterior mean $\mathbb{E}_{\mathcal{Z}_m|\bm{\mathcal{Y}}}[z_m |\mathbf y]$ and the posterior variance $\text{var}_{\mathcal{Z}_m|\bm{\mathcal{Y}}}[z_m|\mathbf y]$. Define $PI^q_m = \int z_m^q \Phi(z_m/\sqrt{v}) \mathcal{N}(z;\widehat{p}_m,\tau^p_m) dz_m$ for  $ q= 0,1,$ and $2$. Using the definition of $p_{\mathcal{Y}_m|\mathcal{Z}_m} (y_m|z_m)$ from (\ref{ygivenz1}), and (\ref{ygivenz2}), the normalization constant can be derived as
	\begin{align*}
	\begin{split}
	& Z^p_m = \int p_{\mathcal{Y}_m| \mathcal{ Z}_m} (y_m|z_m) \mathcal{N}_{ \mathcal{ Z}_m} (z_m; \widehat{p}_m, \tau_m^p) dz \\
	&=\gamma \Big(  PI^0_m \delta_m^- + \Big(1 -   PI^0_m\Big) \delta_m^+ \Big) \\ & \quad \quad \quad \quad \quad + (1-\gamma)\Big(\Big(1 -  PI^0_m\Big) \delta_m^- + PI^0_m \delta_m^+ \Big).
	\end{split} 
	\end{align*}
	Next, we evaluate the posterior mean of  $z_m$ as
	\begin{align*}
	\begin{split}
	&\mathbb{E}_{\mathcal{ Z}_m|\bm{\mathcal{Y}}}[z_m|\mathbf{ y}; \widehat{p}_m, \tau_m^p]  = \Big[ \gamma\Big(PI^1_m \delta_m^- + (\widehat{p}_m - PI^1_m) \delta_m^+ \Big) 
	\\& \quad \quad + (1-\gamma) \Big((\widehat{p}_m - PI^1_m) \delta_m^- + PI^1_m \delta_m^+ \Big) \Big] \frac{1}{Z^p_m}. 
	%
	\end{split}
	\end{align*}
	Similarly, we can evaluate $\mathbb{E}_{\mathcal{Z}_m|  \mathcal{Y}_m}[z_m^2|y_m]$ as,
	\begin{align*}
	\begin{split}
	&\mathbb{E}_{\mathcal{ Z}_m|  \bm{\mathcal{Y}}}[z_m^2|\mathbf y; \widehat{p}_m, \tau_m^p]  =  \Big[ \gamma\Big(PI^2_m \delta_m^-  + (\widehat{p}_m^2 + \tau_m^p - PI_m^2) \delta_m^+ \Big) \\& + (1-\gamma) \Big((\widehat{p}_m^2 + \tau_m^p - PI_m^2) \delta_m^- + I_m^2 \delta_m^+ \Big) \Big] \frac{1}{Z^p_m}. 
	\end{split}
	\end{align*}
	For  the evaluation of $Z^p_m$, $\mathbb{E}_{\mathcal{Z}_m|\bm{\mathcal{Y}}}[z_m|\mathbf y;\widehat{p}_m,\tau^p_m]$, and $\mathbb{E}_{\mathcal Z| \bm{ \mathcal{Y}}}[z_m^2|\mathbf y;\widehat{p}_m,\tau^p_m] $, we need to evaluate integrals $PI_m^0, PI_m^1,$ and $PI_m^2$. Integrals $PI_m^q$ for $q = 0,1, \text{and } 2$ can be evaluated in closed-form as 
	\begin{align*}
	\begin{split}
	& PI_m^0 = \Phi\Big( \frac{\widehat{p}_m}{\sqrt{v + \tau_m^p}}\Big),  \\
	& PI_m^1 = 
	\widehat{p}_m PI_m^0 + \frac{ \tau_m^p ~\mathcal{N}(\frac{\widehat{p}_m}{\sqrt{v + \tau_m^p}})}{\sqrt{v + \tau_m^p}}, \\
	& PI_m^2 =
	\tau_m^p~ PI_m^0 + \widehat{p}_m~ PI_m^1+\frac{\tau_m^p~ \widehat{p}_m~ v  \mathcal{N}(\frac{\widehat{p}_m}{\sqrt{v+\tau_m^p}})}{(v+\tau_m^p)^{1.5}}.
	\end{split}
	\end{align*}
	The derivations of the closed-form expressions of the integrals are provided in Appendix A. 
	The posterior variance can be computed as $\text{var}_{\mathcal{ Z}_m|  \bm{\mathcal{Y}}}[z_m| \mathbf y] = \mathbb{E}_{\mathcal{ Z}_m|  \bm{\mathcal{Y}}}[z_m^2|\mathbf y] - (\mathbb{E}_{\mathcal{ Z}_m|  \bm{\mathcal{Y}}}[z_m|\mathbf y])^2$. With $\mathbb{E}_{\mathcal{ Z}_m|  \bm{\mathcal{Y}}}[z_m|\mathbf y]\text{ and } \text{var}_{\mathcal{ Z}_m|  \bm{\mathcal{Y}}}[z_m| \mathbf y], $ non-linear functions $F_1(\cdot)$, and $F_2(\cdot)$ can be computed as defined in (11). 
	
	Next, we derive the analytical expressions for $G_1(\cdot)$ and $G_2(\cdot)$, i.e.,  expressions for $\mathbb{E}_{\mathcal{X}_n|\bm{\mathcal{Y}}}[x_n| \mathbf y;\widehat{r}_n, \tau^r_{n}]$ and $\text{var}_{\mathcal{X}_n|\bm{\mathcal{Y}}}[x_n| \mathbf y;\widehat{r}_n,  \tau^r_{n}]$. The expectation is carried out with respect to the random variable $\mathcal X_n$ given $\widehat{\mathcal R}_n = \widehat{r}_n$ for random variables
	\begin{equation*}
	\widehat{\mathcal R}_n = \mathcal X_n +\mathcal V_n,
	\end{equation*}
	where $\mathcal V_n \sim \mathcal{ N}(0, \tau^r_n)$ and $\mathcal X_n \sim p_{\mathcal{X}_n}(x_n)$ are independent. Therefore, the marginal posterior distribution can be approximated as 
	\begin{eqnarray}\label{psoterior_prob}
	\!\!\!\!\!\!\!\!\!\!p_{\mathcal{X}_n|\bm{\mathcal{Y}}}(x_n|\mathbf{y};\widehat{r}_n,\tau^r_{n})&\!\!=\!\!&\frac{p_{\mathcal{X}}(x_n)\mathcal{N}(x_n;\widehat{r}_n,\tau^r_{n})}{\int_{x_n}p_{\mathcal{X}_n}(x_n)\mathcal{N}(x_n;\widehat{r}_n,\tau^r_{n})}.
	\end{eqnarray}
	For Bernoulli-Gaussian distribution, the first-order moment can be computed as 
	\begin{align*}
	\mathbb{E}_{\mathcal{X}_n|\bm{\mathcal{Y}}}[x_n|\mathbf y;\widehat{r}_n, \tau^r_{n}] &= \frac{1}{Z_n^r}\int x_n~ p_{\mathcal{X}_n|\bm{\mathcal{Y}}}(x_n|\mathbf{y};\widehat{r}_n,\tau^r_n)  dx_n. 
	\end{align*}
	Using (\ref{signalModel}) and (\ref{psoterior_prob}), and some algebra, we can show that the approximate posterior mean can be expressed as 	
	\begin{align}\label{posteriorMean}
	\mathbb{E}_{\mathcal{X}_n|\bm{\mathcal{Y}}}[x_n|\mathbf{y};\widehat{r}_n, \tau^r_{n}] = Z_n^\prime \exp(-\frac{\widehat{r}_n^2}{2(v_x + \tau^r_{n})}) \widehat{r}_n,
	\end{align}
	where $Z_n^\prime = \frac{1}{Z_n^r}\frac{\lambda}{\sqrt{2 \pi}} \frac{v_x}{(v_x + \tau^r_n)^{1.5}}, \text{ and } Z_n^r$ is the normalization constant which is evaluated as
	\begin{align*}
	\begin{split}
	&Z_n^r = \int x_n~ p_{\mathcal{X}_n|\bm{\mathcal{Y}}}(x_n|\mathbf{y};\widehat{r}_n,\tau^r_{n})  dx_n \\ 
	&= \frac{1-\lambda}{\sqrt{2 \pi \tau^r_{n}}} \exp\Big(\frac{-\widehat{r}_n^2}{2\tau^r_{n}}\Big) + \frac{\lambda}{\sqrt{2 \pi (v_x +\tau^r_{n})}} \exp\Big(\frac{-\widehat{r}_n^2}{2(\tau^r_{n} + v_x)}\Big).
	\end{split}
	\end{align*}
	Similarly, we can evaluate the second-order moment as 
	\begin{align}\label{posteriorVariance}
	& \mathbb{E}_{\mathcal{X}_n|\bm{\mathcal{Y}}}[x_n^2|\mathbf{y};\widehat{r}_n, \tau^r_{n}]=	Z_n^\prime \exp(\frac{-\widehat{r}_n^2}{2(v_x + \tau^r_{n})}) (\frac{\widehat{r}_n^2 v_x}{v_x + \tau^r_{n}} + \tau^r_{n}).
	\end{align}

	Using (\ref{posteriorMean}) and (\ref{posteriorVariance}), the non-linear functions $G_1(.)$ and $G_2(.)$ in (\ref{EstUp}) can be evaluated and hence we can carry out the update in (\ref{nonlinear2}) of \texttt{Noisy1bG}. Thus, we have derived all the statistical quantities required to implement one-bit CS with pre- and post-quantization noise. Accounting for the noise leads to an improved signal reconstruction performance. However, we emphasize that there are applications where the receiver has access to SI which can be used to further improve signal reconstruction performance. In the next section, we look into how we can model SI in the sparse signal reconstruction problem and exploit it for better reconstruction performance.
	
	\section{Noisy one-bit CS with SI}
	In this section, we study the problem of signal reconstruction from noisy one-bit compressed measurements when the receiver has access to SI, $\widetilde{ \mathbf{x}}$, which has both support and amplitude information. We design a \texttt{GAMP} based sparse signal reconstruction algorithm taking SI into account. We assume that the SI is erroneous. The error in SI can either be in the amplitude or in the support set of the signal. We assume that the signal has a small fraction of support that is not in the support set of the SI. These errors are random, and hence, we model side information as a noisy version of the signal, i.e.,   
	\begin{align} \label{sideinformation_model}
	\widetilde{\mathcal X}_n = \mathcal{ X}_n +  \mathcal{ V}_n,    \quad \quad             n = 1, 2, \cdots, N
	\end{align}
	where $\mathcal V_n$ is an additive noise. Note that the magnitude of noise $v_n$ for $n \in \{n^\prime: x_{n^\prime} \neq 0 \text{ and } \widetilde{x}_{n^\prime} \neq 0\} \cup \{n^\prime: x_{n^\prime} = 0 \text{ and } \widetilde{x}_{n^\prime} = 0\}$ is relatively small and close to zero. 
	But for the indices $n \in \{n^\prime: x_{n^\prime} \neq 0 \text{ and } \widetilde{x}_{n^\prime} = 0\} \cup \{n^\prime: x_{n^\prime} = 0 \text{ and } \widetilde{x}_{n^\prime} \neq 0\}$, the magnitude of $v_n$ is quite large. This nature of the error vector suggests that only a small fraction of the error vector has significant values, while most of them are close to zero. Since the noise vector is sparse, we model the noise distribution in ($\ref{sideinformation_model}$) by a Laplace distribution as it forces most of its coefficients to be very small, allowing some occasional large values, i.e., it promotes sparsity on the noise vector  \cite{LBCS}. We then use Gaussian distribution to model the noise distribution and develop algorithms for both of these two cases. Through numerical simulations, we will study the gain in reconstruction performances by the algorithms when the noise, $\mathcal V_n$, is modeled by the sparsity promoting distribution, i.e., Laplace distribution. 
	
	\begin{algorithm}

		\begin{enumerate}
			\item \textbf{Initialization}: Set t=0 and initialize $\widehat{\mathbf{x}}^t , \mathbf{\tau}_x^t,\widehat{ \mathbf{s}}^t$ and $ \mathbf v_s$ as $\widehat{\mathbf{x}}^t = \mathbb{E}[\mathbf{x}],~~  \mathbf{\tau}_x^t = \text{var}[\mathbf{x}], ~~\widehat{ \mathbf{s}}^t = {0}, 
			$ and $\mathbf v_s = {0}$ where the expectation and variance of $\mathbf{x}$ are with respect to $p_x$
			\item \textbf{While} loop $ l < L $
			\item \textbf{While} loop $t<T$
			\item \textbf{Measurement Update}
			\begin{itemize}
				\item[] Same as in Algorithm1 
			\end{itemize}
			
			\item \textbf{Estimation Update} 
			\begin{itemize}
				\item[] Linear Step
				{ \small 	\begin{align*}
					&\boldsymbol\tau^{r,t+1} = 			((\mathbf A \odot \mathbf A)^T \boldsymbol{\tau}_s^t)^{-1}, 
					\\& 
					\widehat{ \mathbf{r}}^{t+1} =  \widehat{ \mathbf{x}}^{t} + \boldsymbol{\tau}^{r,t} \odot(\mathbf A^T  \widehat{ \mathbf{s}}^{t+1}),		 	     \end{align*} }	
				where the inversion is performed element-wise		
				\item[] Non-linear Step
				{ \small 	\begin{align}
					&\widehat{ \mathbf{x}}^{t+1}  =  G_1( \widehat{ \mathbf{r}}^{t+1}, \boldsymbol\tau^{r,t+1};p_{\mathcal{X}| \bm{\mathcal{Y}}, \widetilde{\mathcal{X}}})\\ 
					&{ \boldsymbol{\tau}}^{t+1}_x =    G_2( \widehat{ \mathbf{r}}^{t+1}, \boldsymbol\tau^{r,t+1};p_{\mathcal{X}| \bm{\mathcal{Y}}, \widetilde{\mathcal{X}}}),	 	   
					\end{align} }
				where $ G_1$ and $ G_2$ are applied element-wise and are defined as 
				{ \small \begin{align}
					\begin{split}
					&G_1( \widehat{r}_n, \tau^r_n;p_{\mathcal{X}| \bm{\mathcal{Y}}, \widetilde{\mathcal{X}}})= \mathbb{E}_{\mathcal{X}_n|\bm{\mathcal{Y}},\widetilde{X}_n}[x_n|\mathbf y, \widetilde{x}_n;\widehat{r}_n,\tau^r_{n}], \\&
					G_2( \widehat{r}_n,\tau^r_n;p_{\mathcal{X}| \bm{\mathcal{Y}}, \widetilde{\mathcal{X}}}) = \text{var}_{\mathcal{X}_n|\bm{\mathcal{Y}},\widetilde{X}_n}[x_n|\mathbf y, \widetilde{x}_n;\widehat{r}_n,\tau^r_{n}].
					\end{split}
					\end{align} }
			\end{itemize}
			The expectation and variance are evaluated with respect to $p_{ \mathcal{X}_n|\widetilde{\mathcal{X}}_n, \bm{\mathcal{Y}}} \propto \mathcal{N}(\cdot; \widehat{r}_n, \tau^r_n) p_{\mathcal X}(\cdot) p_{\widetilde{\mathcal{X}}_n|\mathcal{X}_n}(\cdot),$ and can be computed by using (\ref{SI_Eq})  
			\item []  Set t = t+1 
			\item \textbf{End While}
			\item Update $v_s$ using (\ref{varPar})
			\item $l = l+1$
			\item \textbf{End While}
			\caption{\texttt{GAMP} Algorithm for noisy  one-bit CS with SI}\label{dalgo2}
		\end{enumerate}
		
	\end{algorithm}
	
	\vspace{-.05 cm}
	\subsection{Noisy one-bit CS with Laplacian Noise (\texttt{laplacianSI})}
	In this subsection, we model the noise in SI as a Laplacian distributed random variable.
	Thus, we choose  $p_{\bm{\widetilde{\mathcal{X}}}|\bm{\mathcal{X}}}(\widetilde{\mathbf{x}}|\mathbf{x})$ as
	\begin{equation}
	p_{\bm{\widetilde{\mathcal{X}}}|\bm{\mathcal{X}}}(\widetilde{\mathbf{x}}|\mathbf{x}) = \Big(\frac{1}{4v_s}\Big)^N \exp(- \frac{\| \mathbf{x}- \widetilde{\mathbf{x}}\|_1}{2v_s}),
	\end{equation}
	where $v_s$ is a constant that determines the variance of the distribution and it captures the confidence that the receiver has on how close SI is to the sparse signal.
	
	\begin{table*}
		\caption{\texttt{GAMP} Equations for Side-Information}
		\centering
		\vspace{- .45 cm}
		\begin{minipage}{1.0\textwidth}
			\begin{align} 
			\begin{split}
			&Z^l_n =\frac{1- \lambda}{4v_s\sqrt{2\pi \tau^r_n}} \exp(-\frac{\widehat{r}_n^2}{2\tau^r_n} - \frac{|\widetilde{x}_n|}{2 v_s}) 
			+ \lambda \phi\big(\frac{\widehat{r}_n}{\sqrt{\tau^r_n+v_x}}\big)
			\Big(C_{1,n} \Phi(\frac{m^l_n- (m^g_n+ \frac{v^g_n}{2v^l_n})}{\sqrt{v^g_n}}) + C_{2,n} \Bigg(1-\Phi(\frac{m^l_n- (m^g_n- \frac{v^g_n}{2v^l_n})}{\sqrt{v^g_n}}) \Bigg)
			\\&
			\mathbb{E}_{\mathcal{X}_n|\bm{\mathcal{Y}},\widetilde{\mathcal X}_n}[x_n|\mathbf y, \widetilde{x}_n;\widehat{r}_n,\tau^r_{n}] = \frac{\lambda \phi\big(\frac{\widehat{r}_n}{\sqrt{\tau^r_n+v_x}}\big)}{Z^l_n}  \Bigg(C_{1,n} I_1(m^l_n; m^g_n +\frac{v^g_n }{2v^l_n }, v^g_n ) +  C_{2,n} \Big(m^g_n -\frac{v^g_n }{2v^l_n } - I_1(m^l_n ; m^g_n-\frac{v^g_n}{2v^l_n}, v^g_n ) \Big) \Bigg)
			\\&
			\mathbb{E}_{\mathcal{X}_n|\bm{\mathcal{Y}},\widetilde{\mathcal X}}[x_n^2|\mathbf y, \widetilde{x}_n;\widehat{r}_n,\tau^r_{n}] = \frac{\lambda \phi\big(\frac{\widehat{r}_n}{\sqrt{\tau^r_n+v_x}}\big)}{Z^l_n} \Bigg(C_{1,n} I_2\Big(m^l_n; m^g_n +\frac{v^g_n }{2v^l_n }, v^g_n \Big) + 
			C_{2,n} \Big( \Big(m^g_n -\frac{v^g_n }{2v^l_n }\Big)^2+ v^g_n - I_2\Big(m^l_n; m^g_n -\frac{v^g_n }{2v_l}, v^g_n \Big) \Big) \Bigg).
			\end{split} \label{SI_Eq}
			\end{align}
			\medskip
			\hrule
		\end{minipage}
	\end{table*}
	
	Next, we develop a \texttt{GAMP}-based algorithm for one-bit CS with side-information. Note that the evaluation of $\mathbb{E}_{\mathcal Z_m|  \bm{\mathcal{Y}}}[z_m|\mathbf y]$ and  $\mathbb{E}_{\mathcal{ Z}_m|  \bm{\mathcal{Y}}}[z_m^2| \mathbf y]$ depends only on the distribution of the channel and hence is the same as in Algorithm 1. Next, we derive expressions for $G_1(\cdot)$ and  $G_2(\cdot)$ when the receiver has access to SI. Here, we assume that the noise is Laplacian.
	The expectation is carried out with respect to random variable $\mathcal X_n$ given $\widehat{ \mathcal R}_n = \widehat{r}_n,$ and $\widetilde{ \mathcal X}_n = \widetilde{x}_n$ for random variables
	\begin{equation*} 
	\widehat{\mathcal R}_n = \mathcal{X}_n + \mathcal{V}_n, ~~~~~~~~~    \widetilde{\mathcal{ X}}_n = \mathcal{ X}_n + \mathcal{ W}_n, 
	\end{equation*}
	where $\mathcal{ V}_n \sim \mathcal{N}(0, \tau^n_r)$,  $\mathcal{W} \sim \mathcal{L}(0,2v_s)$ and $\mathcal{X}_n \sim p_{\mathcal{X}_n}(x_n)$ are independent. Therefore, the marginal posterior distribution can be approximated as 
	
	\begin{align*}\label{psoterior_prob_SI1}
	p_{\mathcal{X}_n|\bm{\mathcal{Y}}}(x_n|\mathbf{y};\widehat{r}_n,\tau^r_{n})=\frac{p_{\mathcal{X}_n}(x_n)\mathcal{N}(x_n;\widehat{r}_n,\tau^r_{n}) \mathcal{L}(x_n;\widetilde{x}_n, 2v_s)}{\int_{x_n}p_{\mathcal{X}_n}(x_n)\mathcal{N}(x;\widehat{r}_n,\tau^r_{n}) \mathcal{L}(x_n;\widetilde{x}_n, 2v_s)}.
	\end{align*}
	Using the approximated posterior density function, $p_{\mathcal{X}_n|\bm{\mathcal{Y}},\widetilde{X}_n}(x_n|\mathbf{y},\widetilde{x}_n,\widehat{r}_n,\tau^r_{n})$, we evaluate the first-order moment, $\mathbb{E}_{\mathcal{X}_n|\bm{\mathcal{Y}},\widetilde{X}_n}[x_n|\mathbf y, \widetilde{x}_n;\widehat{r}_n,\tau^r_{n}]$ and second-order moment, $\mathbb{E}_{\mathcal{X}_n|\bm{\mathcal{Y}};\widetilde{X}}[x_n^2|\mathbf y, \widetilde{x}_n;\widehat{r}_n,\tau^r_{n}]$. 
	\begin{result} \label{Result1}
		Define $m^g_n \triangleq \frac{v_x \widehat{r}_n}{v_x+\tau^r_n}$, $v^g_n  \triangleq \frac{v_x \tau^r_n}{v_x+\tau^r_n}, m^l_n \triangleq \widetilde{x}_n, v^l_n \triangleq v_s, C_{1,n} \triangleq \frac{1}{4 v^l_n} \exp(-\frac{1}{2v^l_n}({m^l_n-m^g_n-\frac{v^g_n}{4v^l_n}}) $, and $C_{2,n} \triangleq   \frac{1}{4 v^l_n} \exp(-\frac{1}{2v^l_n}\big(-m^l_n+m^g_n-\frac{v^g_n}{4v^l_n}\big))$. The posterior first-order and second-order moments are listed in (\ref{SI_Eq}).
	\end{result}
	
	The sketch of the derivations is provided in Appendix B. The first-order and second-order moments require evaluation of integrals $I_0(\cdot), I_1(\cdot)$ and $I_2(\cdot)$. We have the following results on the closed-form expressions of these integrals. 
	
	
	\begin{result}
		With $I_q(\widetilde{x}_n; m_n,\tau^r_n) \triangleq \int_{-\infty}^{\widetilde{x}_n} x_n^q \mathcal{N}(x_n|m_n,\tau^r_n) dx_n$, the analytical expressions of $I_n^1$ and $I_n^2$ are
		\begin{align}
		\begin{split}
		&I_1( \widetilde{x}_n; m_n, \tau_n^r) = 
		m_n \Phi(\frac{\widetilde{x}_n - m_n}{\sqrt{\tau^r_n}}) - \sqrt{\tau^r_n }\phi(\frac{\widetilde{x}_n-m_n}{\sqrt{\tau^r_n}}) \\
		&I_2( \widetilde{x}_n; m_n, \tau_n^r)  = m_n I_1(\widetilde{x}_n; m_n, \tau^r_n) + \tau^r_n ~ \Phi(\frac{\widetilde{x}_n-m_n}{\sqrt{\tau^r_n}}) 
		\\& ~~~~~~~~~~~~~~~~~~~~~~~~ - \widetilde{x}_n \sqrt{\tau^r_n} \phi(\frac{\widetilde{x}_n-m_n}{\sqrt{\tau^r_n}}).
		\end{split}
		\end{align}
	\end{result} 
	The sketch of the proofs of Result 2 is provided in Appendix C. 
	With posterior first-order moment and second-order moments, we have all the statistical quantities required to implement Algorithm 2. 
	\subsection{Estimation of the $v_s$}
	In the following, we employ the Expectation-Maximization (EM) algorithm to estimate the side-information parameter, $v_s$. The EM algorithm is an iterative technique that increases the lower bound on the likelihood $p(\mathbf y; v_s)$ at each iteration, which guarantees that the likelihood converges to a local maximum, or at least to a saddle point. Specifically, the EM algorithms iterates over two steps: 1) \textit{Expectation step}: choosing distribution to maximize the lower bound for fixed $v_s = v_s^k$ , and 2) \textit{Maximization step}: choosing $v_s$ to maximize the lower bound for the fixed distribution from Step 1. We emphasize that the maximizing pdf  is the true posterior under the prior parameter, $v_s$. Since, it is very difficult to compute the true posterior, we use the posterior approximated by the \texttt{GAMP} algorithm in the evaluation of the expectation.  
	The EM algorithm is summarized as 
	\begin{equation} \label{EM_problem}
	v_s^{k+1} = \underset{v_s}{ \text{argmin }} ~\mathbb{E}_{\bm{\mathcal{X}}|\bm{\mathcal{Y}}, \bm{\widetilde{\mathcal{X}}}; v_s^k}[- \log p(\mathbf y, \mathbf x,\widetilde{ \mathbf x}; v_s)],
	\end{equation}
	where $p(\mathbf x, \mathbf y,  \widetilde{ \mathbf  x}, v_s)$ is the joint probability distribution of the complete data and $p(\mathbf{x}| \mathbf y,  \widetilde{ \mathbf x}, v_s^k)$ is the approximated posterior density given the side-information which is parameterized by the previous iteration estimate of $v_s^k$. We first carry out the expectation step as
	\begin{align*}
	\begin{split}
	&\mathbb{E}_{\bm{\mathcal{X}}|\bm{\mathcal{Y}}, \bm{\widetilde{\mathcal{X}}}; v_s} [-\log ~p(\mathbf y, \mathbf x, \widetilde{ \mathbf  x}, v_s)] 
	= \\
	& \mathbb{E}_{\bm{\mathcal{X}}|\bm{\mathcal{Y}}, \bm{\widetilde{\mathcal{X}}}; v_s}[-\log ~ p(\mathbf y| \mathbf x, \widetilde{ \mathbf x}, v_s^t)  - \log p(\mathbf x , \widetilde{ \mathbf  x}|  v_s) - \log p(v_s) ].
	\end{split}
	\end{align*}
	We note that, the expectation step is followed by the maximization step, and all the terms that do not involve $v_s$ eventually go to zero. Since $\log ~ p(\mathbf y| \mathbf x, \widetilde{ \mathbf x})$ does not depend on $v_s$, we drop the term. Similarly, we will drop all the terms that do not depend on $v_s$ in the subsequent steps.
	\begin{align*}
	\centering
	\begin{split}
	& \mathbb{E}_{\bm{\mathcal{X}}|\bm{\mathcal{Y}}, \bm{\widetilde{\mathcal{X}}}; v_s}[- \log p(\mathbf x , \widetilde{ \mathbf  x}| v_s) - \log p(v_{s}) ]
	= \\
	&  \sum_{n = 1}^{N}\Bigg( \mathbb{E}_{{\mathcal{X}}_n|\bm{\mathcal{Y}}, \bm{\widetilde{\mathcal{X}}}; v_s} \Bigg( \frac{|x_n-\widetilde{x}_n|}{2v_{s}} \Bigg) + \log(v_s) - \log p(v_{s})\Bigg),
	\end{split}
	\end{align*}
	where the summation over indices is due to the fact that the posterior density, $p_{\bm{\mathcal{X}}|\bm{\mathcal{Y}}}$ is approximated as  $p_{\bm{\mathcal{X}}|\bm{\mathcal{Y}}} = \prod_{n =1}^{N} p_{\mathcal{X}_n|\bm{\mathcal{Y}}}$.
	From (\ref{EM_problem}), the estimation of $v_s$ can be written as
	\begin{align*}
	\begin{split}
	& v_s^{k+1}=
	\\&
	\underset{v_s}{ \text{argmin }} \sum_{n = 1}^{N} \mathbb{E}_{{\mathcal{X}}_n|\bm{\mathcal{Y}}, \bm{\widetilde{\mathcal{X}}}; v_s^k}  \frac{|x_n-\widetilde{x}_n|}{2v_{s}}  + \log(v_{s}) - \log p(v_{s}).
	\end{split}
	\end{align*}
	We assume a non-informative prior on the parameter $v_{s}$. Hence, we drop the $\log p(v_{s})$ term and find the maximum likelihood estimate of $v_{s}$ as
	\begin{align}
	\begin{split}
	&v_{s}^{k+1} = \frac{1}{2N}~~ \sum_{n = 1}^{N} \mathbb{E}_{{\mathcal{X}}_n|\bm{\mathcal{Y}}, \bm{\widetilde{\mathcal{X}}}; v_s^k} ~ |x_n-\widetilde{x}_n| 	\end{split} \label{EPvarSILap}
	\end{align}
	With the notations as defined in Result \ref{Result1}, we can evaluate the expectation in (\ref{EPvarSILap}) in closed-form as 
	\begin{align}
	\begin{split}
	&\mathbb{E}_{{\mathcal{X}}_n|\bm{\mathcal{Y}}, \bm{\widetilde{\mathcal{X}}}; v_s^k}  ~ |x_n-\widetilde{x}_n| = \\ 
	& - \int_{- \infty}^{\widetilde{x}} (x_n - \widetilde{x}_n) \mathcal{N}(x_n;\widehat{r}_n, \tau^r_n) p_{\mathcal{X}_n}(x_n) \exp(\frac{x_n- \widetilde{x}_n}{2v^k_s}) dx_n   
	\\
	& + \int_{\widetilde{x}_n}^{\infty} (x_n - \widetilde{x}_n) \mathcal{N}(x_n;\widehat{r}_n, \tau^r_{n}) p_{\mathcal{X}_n}(x_n) \exp(-\frac{x_n- \widetilde{x}_n}{2v^k_s}) dx_n 
	\\ & = \frac{ 1}{Z^l_n} \Big(  \lambda \phi\big(\frac{\widehat{r}_n}{\sqrt{\tau^r_{n}+v_x}}\big) C_{2,n} \Big(m^g_n -\frac{v^g_n }{2v^l_n } - I_1(m^l_n ; m^\prime_G-\frac{v^g_n}{2v^l_n}, v^g_n) 
	\\&  -\widetilde{x}_n ( 1 - \Phi(\frac{m^l_n - (m^g_n - \frac{v^g_n}{2v^l_n})}{\sqrt{v^g_n}}) )  \Big)
	- \lambda C_{1,n} \phi\big(\frac{\widehat{r}_n}{\sqrt{\tau^r_{n}+v_x}}\big)  \\& \quad \quad \quad \Big(I_1(m^l_n; m^g_n +\frac{v^g_n }{2v^l_n }, v^g_n )   - \widetilde{x}_n \Phi(\frac{m^l_n - (m^g_n + \frac{v^g_n}{2v^l_n})}{\sqrt{v^g_n}})\Big)
	\\& \quad \quad \quad \quad \quad \quad \quad \quad  +  |\widetilde{x}_n| (1-\lambda) \mathcal{N}(0;\widehat{r}_n,\tau^r_n) \mathcal{L}(0;\widetilde{x}_n,2v^k_s) \Big)
	\end{split}\label{varPar}
	\end{align}
	Using (\ref{varPar}) in (\ref{EPvarSILap}), we find the estimate of the $v_s$ using the EM algorithm.  
	
	Hence, we have derived all the expressions required for signal reconstruction from one-bit measurements with side-information.  
	In Algorithm 2, we summarize the steps for signal reconstruction for one-bit compressed sensing with side-information with parameter estimation.

	\subsection{Noisy one-bit CS with Gaussian Noise (GaussianSI)}
	Next, we list the steps for the estimation of the sparse signals when the side-information is assumed to be the actual signal corrupted by Gaussian noise.
	\begin{equation*}
	\widehat{\mathcal R}_n = \mathcal{X}_n + \mathcal{V}_n, ~~~~~~~~~    \widetilde{\mathcal X}_n = \mathcal{X}_n + \mathcal{W}_n, 
	\end{equation*}
	where $\mathcal{V}_n \sim \mathcal{N}(0, \tau^r_{n})$,  $\mathcal{W}_n \sim \mathcal{N}(0,v_s)$ and $\mathcal{X}_n \sim p_{\mathcal{X}_n}(x_n)$ are independent. Next, we state the results for the first and second order moments for this setup. 
	\begin{result}
		The posterior first-order and second-order moments of the signal given side-information, $\widetilde{ x}_n$ are
		\begin{align}
		\begin{split} \label{UpdateEqn_Gaussian}
		&\mathbb{E}_{\mathcal{X}_n|\bm{\mathcal{Y}},\widetilde{X}_n}[x_n|\mathbf y, \widetilde{x}_n;\widehat{r}_n,\tau^r_{n}] = \pi^g_{n} \frac{\widehat{r}_n v_s v_x + v_x \tau^r_{n} \widetilde{x}}{v_x\tau^r_{n} + \tau^r_{n}v_s + v_s v_x} \triangleq \pi^g_n m^g_n\\
		&\mathbb{E}_{\mathcal{X}_n|\bm{\mathcal{Y}},\widetilde{X}_n}[x^2_n|\mathbf y, \widetilde{x}_n;\widehat{r}_n,\tau^r_{n}] = \pi^g_{n} \Big(\frac{v_s \tau^r_{n} v_x}{v_x\tau^r_{n} + \tau^r_{n}v_s + v_s v_x} + \big(m^g_n\big)^2\big),
		\end{split}
		\end{align} 
	\end{result}
	where $\pi^g_n = \frac{ \lambda }{\lambda + (1-\lambda) Z_n} $ and
	$Z_n = \frac{\mathcal{N}(0;\widehat{x}, \tau^r_{n}) \mathcal{N}(0;\widetilde{x}_n, v_s)}{\mathcal{N}(0;\widehat{r}_n, v_x + \tau^r_{n}) \mathcal{N}(0;\frac{\widehat{r}_n v_x}{v_x+\tau^r_{n}}-\widetilde{x}, \frac{\tau^r_{n}v_x}{v_x + \tau^r_{n}} + v_s)}.$
	The sketch of derivations is provided in Appendix D.  Next, we estimate the side-information parameter, $v_s$ using the EM algorithm. Following the steps as in the Laplacian noise case, we can show that the maximum likelihood estimator of the $v_s$ is, 
	\begin{align}
	\begin{split}
	v_{s}^{k+1} &= \frac{1}{N}~~ \sum_{n = 1}^{N} \mathbb{E}_{\mathcal{X}_n|\bm{\mathcal{Y}}, \widetilde{\mathcal{X}}_n;  v_{s}^k} ~ (x_n-\widetilde{x}_n)^2 \\
	& =  \frac{1}{N}~~ \sum_{n = 1}^{N} \mathbb{E}_{\mathcal{X}_n|\bm{\mathcal{Y}}, \widetilde{\mathcal{X}}_n; v_{s}^k} (x_n^2) - 2 \mathbb{E}_{\mathcal{X}_n|\bm{\mathcal{Y}}, \widetilde{\mathcal{X}}_n; v_{s}^k}(x_n) + \widetilde{x}_n^2 \\
	& = \frac{1}{N} \sum_{n = 1}^{N} \pi^g_n \Bigg(  \frac{v^k_s \tau^r_{n} v_x}{v_x\tau^r_{n} + \tau^r_{n}v^k_s + v^k_s v_x} 
	\\& - \Big( \pi^g_n \frac{\widehat{r}_n v^k_s v_x + v_x \tau^r_{n} \widetilde{x}_n}{v_x\tau^r_{n} + \tau^r_{n}v^k_s + v^k_s v_x}\Big)^2 - 2\widetilde{x}_n m^g_n + \widetilde{x}_n^2
	\Bigg)
	\end{split}\label{estimate_side_Gaussian}
	\end{align}
	where the equality is obtained by replacing  $\mathbb{E}_{\mathcal{X}_n|\bm{\mathcal{Y}}, \widetilde{\mathcal{X}}_n;  v_{s}^k} (x_n^2)$ and $\mathbb{E}_{\mathcal{X}_n|\bm{\mathcal{Y}}, \widetilde{\mathcal{X}}_n;  v_{s}^k} (x_n^2)$ from  (\ref{UpdateEqn_Gaussian}).
	
	With (\ref{UpdateEqn_Gaussian}) and (\ref{estimate_side_Gaussian}), we have evaluated all the expressions required for implementing the one-bit compressed sensing algorithm with Gaussian side-information. 
	In the simulation section, we will discuss that modeling noise with Laplacian distribution in noise makes the proposed algorithm more robust when the side-information has partial support information or when the support in the side-information is erroneous .
	
	\section{Noisy one-bit CS with noisy support as side-information}
	In this section, we investigate the problem of sparse signal reconstruction from noisy one-bit compressed measurements when the receiver has access to only support-information as SI. We develop a \texttt{GAMP} based algorithm by taking support information as SI into account. We assume that there are some discrepancies between the support of the signal and the SI. We model these discrepancies using multiplicative noise.
	Formally, let $\mathcal{S}_n$ be a random variable that represents the support of the $n$-th index of the sparse signal, $\mathcal{X}_n$, for $n = 1, \cdots, N$. $\mathcal{S}_n$ takes values $1$ and $-1$ depending on whether or not the signal index is in the support set of $\mathcal{X}$, i.e.,
	\begin{eqnarray}
	s_n = \left\{
	\begin{array}{ccc}
	+1, ~&~ \textrm{if}~ x_{n} \neq 0,\\
	-1, ~&\textrm{if}~ x_{n}  = 0,\\
	\end{array}\right.\label{support_encoding}
	\end{eqnarray}
	Let $\widetilde{ \mathcal{X}}_n$ be the $n$-th element of SI which is the noisy version of the actual support of the signal, i.e., $\mathcal{S}_n$. We assume that a small fraction of the support set is different (erroneous) in the SI from that of the signal. We model this relationship between SI and the actual support-set of the signal by
	\begin{equation*}
	\widetilde{\mathcal{X}}_n = \zeta_n \mathcal{S}_n, 
	\end{equation*}
	where $\zeta_n$ is the multiplicative noise which can take values $1$ or $-1$, and is assumed to be a Bernoulli distributed random variable with probability $\beta$ for event $\zeta_n$ = 1 and probability $1-\beta$ for event $\zeta_n =-1$, respectively. 
	Thus
	\begin{align*}
	& p(\widetilde{ x}_n =1|s_n = -1)  = p(\widetilde{ x}_n =-1|s_n = 1) = 1-\beta, \\
	& p(\widetilde{ x}_n =1|s_n = - 1)  = p(\widetilde{ x}_n = -1|s_n = -1) = \beta
	\end{align*}
	Next, we develop a \texttt{GAMP}-based algorithm for one-bit CS with erroneous support information as SI. As the evaluation of $\mathbb{E}_{\mathcal Z|  \mathcal{Y}}[z|y]$ and  $\mathbb{E}_{\mathcal Z|  \mathcal{Y}}[z^2|y]$ depends only on the distribution of the channel (\ref{channel}), $F_1(\cdot)$ and $F_2(\cdot)$ are essentially the same as in Section III. Next, we derive expressions for
	$\mathbb{E}_{\mathcal{X}_n|\bm{\mathcal{Y}},\widetilde{X}_n}[x_n|\mathbf y, \widetilde{x}_n;\widehat{r}_n,\tau^r_{n}]$ and  $\mathbb{E}_{\mathcal{X}_n|\bm{\mathcal{Y}};\widetilde{X}}[x_n^2|\mathbf y, \widetilde{x}_n;\widehat{r}_n,\tau^r_{n}]$
	when the receiver has access to noisy support information as SI.
	The expectation is carried out with respect to the random variable $\mathcal{X}_n$ given $\widehat{\mathcal{R}}_n = \widehat{r}_n,$ and $\widetilde{ \mathcal{X}}_n = \widetilde{x}_n$ for random variables
	\begin{equation*}
	\widehat{\mathcal{R}}_n = \mathcal{X}_n + \mathcal{V}_n, ~~~~~~~~~    \widetilde{\mathcal{X}}_n = \zeta_n \mathcal{S}_n
	\end{equation*}
	where $\mathcal V_n$,  $\zeta_n$ and $\mathcal{X}_n$ are independent. Therefore, the marginal posterior distribution can be approximated as 
	
	\begin{align}\label{posterior_prob_support}
	p_{\mathcal{X}_n|\bm{\mathcal{Y}}}(x_n|\mathbf{y};\widehat{r}_n,\tau^r_{n})=\frac{p_{\mathcal{X}_n}(x_n)\mathcal{N}(x_n;\widehat{r}_n,\tau^r_{n}) p_{\mathcal{\zeta}_n}(\zeta_n = \frac{\widetilde{x}_n}{s_n})}{\int_{x_n} p_{\mathcal{X}_n}(x_n)\mathcal{N}(x_n;\widehat{r}_n,\tau^r_{n}) p_{\mathcal{\zeta}_n}(\zeta_n = \frac{\widetilde{x}_n}{s_n})}.
	\end{align}
	With  (\ref{posterior_prob_support}) as the approximated marginal posterior density function, we express the analytical expression for posterior first-order and second-order moments as 
	\begin{result}
		Let $\pi_n$ be the posterior probability of $x_n$ being a non-zero element. Then 
		\[ \pi_n =  \frac{\lambda p(\widetilde{x}_n|s_n=1) }{\lambda p(\widetilde{x}_n|s_n=1) + (1-\lambda) p(\widetilde{x}_n|s_n =0) Z_n} \]
		where $Z_n = \frac{\mathcal{N}(0;\widehat{r}_n, \tau^r_n)}{\mathcal{N}(0;\widehat{r}_n, v_x + \tau^r_n)}$.
		The posterior first-order and second-order moments of the sparse signal given noisy support-information, $\widetilde{x}_n$, are
		\begin{align}
		\begin{split} \label{UpdateEqn_binary}
		&\mathbb{E}_{\mathcal{X}_n|\bm{\mathcal{Y}},\widetilde{X}_n}[x_n|\mathbf y, \widetilde{ x}_n; \widehat{r}_n, \tau^r_n] 
		= \pi_n \frac{\widehat{r}_nv_x}{v_x+\tau^r_{n} }
		\\ 
		&\mathbb{E}_{\mathcal{X}_n|\bm{\mathcal{Y}},\widetilde{X}_n}[x^2_n|\mathbf y, \widetilde{ x}_n; \widehat{r}_n, \tau^r_n]  
		= \pi_n \Bigg( \frac{\tau^r_{n} v_x}{v_x + \tau^r_{n}} + \Big( \frac{\widehat{r}_n v_x}{v_x + \tau^r_{n}}\Big)^2 \Bigg),
		\end{split}
		\end{align}
	\end{result}
	\noindent where $Z_n = \frac{\mathcal{N}(0;\widehat{x}, \tau^r_{n})}{\mathcal{N}(0;\widehat{r}_n, v_x + \tau^r_{n})}.$ 
	We can obtain the above results by the substitution of $p_{\mathcal{X}_n}(x_n)$ from (\ref{signalModel}) in (\ref{posterior_prob_support}), followed by representing the posterior density, $p_{\mathcal{X}_n|\bm{\mathcal{Y}}}$,  as a Bernoulli-Gaussian pdf: $(1-\pi_n) \delta(x) + \pi_n \mathcal N(x;m,v)$, and using the definition of first-order and second-order moments. Since the derivation of the first-order and the second-order moments is similar to the case when the noise is assumed to Laplacian, we omit the actual derivations. 
	Note that, we assumed noisy support-information in the problem statement. Next, we estimate the noise parameter using the EM algorithm. Following the EM algorithm based approach in the previous section, the maximum likelihood estimate of $\beta$ is
	\begin{align*}
	\centering
	\begin{split}
	\beta^{t+1} &= \underset{\beta}{ \text{argmin }} ~\mathbb{E}_{\bm{\mathcal{X}}|\bm{\mathcal{Y}}, \bm{\widetilde{\mathcal{X}}}; \beta}[- \log p(\mathbf y, \mathbf x,\widetilde{ \mathbf x}; \beta)]
	\\
	&= \underset{\beta}{ \text{argmin }} \mathbb{E}_{\bm{\mathcal{X}}|\bm{\mathcal{Y}}, \bm{\widetilde{\mathcal{X}}}; \beta}[- \log p( \widetilde{ \mathbf  x}|\mathbf x; \beta)]
	\end{split}
	\end{align*}
	\noindent With $\pi_n$ as the posterior probability of $n$-th element of $\mathbf{x}$ being non-zero, the expectation can be evaluated as 
	\begin{align*}
	\begin{split}
	&\mathbb{E}_{\bm{\mathcal{X}}|\bm{\mathcal{Y}}, \bm{\widetilde{\mathcal{X}}}; \beta}[- \log p( \widetilde{ \mathbf  x}|\mathbf x ; \beta)] = 
	\\&
	\sum_{\{n:\widetilde{x}_n=1\}} \log(1-\beta)(1-\pi_n)  
	+ \log(\beta)\pi_n
	\\&
	+\sum_{\{n:\widetilde{x}_n=-1\}}  \log(1-\beta)\pi_n
	+ \log(\beta) (1-\pi_n),
	\end{split}
	\end{align*} 
	Next, we estimate the value of $\beta$ that maximizes the expectation. Differentiating the expectation with respect to $\beta$ and equating to zero, we get
	\begin{align} \label{EM_binarySI}
	\beta =  \frac{\sum_{\{n:\widetilde{x}_n=1\}}\pi_n  + \sum_{\{n:\widetilde{x}_n=-1\}}(1-\pi_n) }{ N}
	\end{align}
	With the results in  (\ref{UpdateEqn_binary}) and (\ref{EM_binarySI}), we can use Algorithm 2 for estimating sparse signals from their one-bit compressed measurements with erroneous support information as the SI. Next, we provide simulation results for the proposed algorithms.

	\begin{figure*}[htb] 
		\centering
		\begin{subfigure}[b]{0.45\textwidth}
			\includegraphics[width = .9\textwidth]{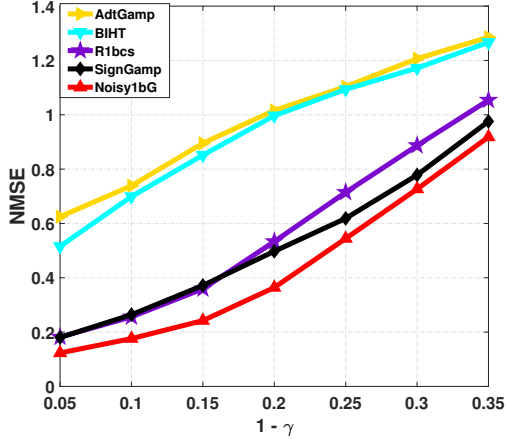}
			\caption{NMSE against bit-flip probability}
			\label{fig:gamma}
		\end{subfigure} 
		~ \begin{subfigure}[b]{0.45\textwidth}
			\includegraphics[width=.9\textwidth]{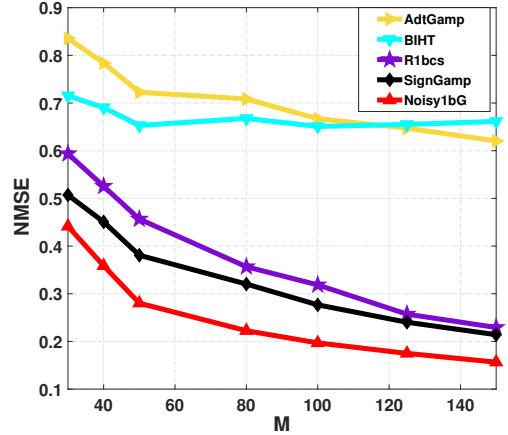}
			\caption{ \footnotesize NMSE as a function of $M$}
			\label{fig:M}
		\end{subfigure} 
		\caption{Comparison of reconstruction performance of the proposed method when $N=50$, $\lambda = 0.15$.} \label{fig:1bitalgorithm}
	\end{figure*}
	\section{Simulation Results}
	In this section, we evaluate the signal reconstruction performance of the proposed sparse signal reconstruction algorithms from noisy one-bit measurements with the state-of-the-art algorithms. We consider the problem of reconstructing a sparse signal of dimension $N$ from $M$ noisy one-bit measurements. The measurement matrix, $\mathbf{A} \in \mathbb{R}^{M \times N}$ is drawn from an i.i.d. Gaussian distribution with zero-mean and unit variance. We consider real-valued compressed measurements that are corrupted by AWGN noise before quantization and the sign-flip noise (Bernoulli) after quantization. We employ normalized mean square error (NMSE) as the  performance metric which is defined as
	\begin{align*}
	\text{NMSE} = \sqrt{\Bigg\|\frac{\mathbf{x}}{\|\mathbf{x} \|}-\frac{\widehat{\mathbf{x}}}{\|\widehat{\mathbf{x}}\|} \Bigg\|^2_2}   
	\end{align*}
	where $\mathbf x$ and $\mathbf{\widehat{ x}}$ are the actual signal and the reconstructed signal, respectively. We generate a sparse signal vector from the Bernoulli-Gaussian distribution with signal sparsity parameter $\lambda = 0.1$, mean zero and variance 5.5. We assume that the signal is corrupted by additive white Gaussian noise before quantization with mean zero and covariance $v\mathbb{I}_N$. After quantization, the one-bit quantized measurements are corrupted by sign-flip noise generated from Bernoulli distribution with probability of sign flip $1-\gamma.$ We ran the algorithm for 500 Monte-Carlo runs.

	In the first experiment, we evaluate the performance of the proposed one-bit CS algorithm with the state-of-the-art algorithms.  In this experiment, we compare the performance of the proposed algorithm, \texttt{Noisy1bG}, with algorithms proposed in \cite{Jacques1}, \cite{kamilov}, and \cite{r1bcs} respectively and refer to these algorithms as \texttt{BIHT}, \texttt{AdtGamp}, and \texttt{R1bcs}.  \texttt{SignGAMP} refers to the one-bit \texttt{GAMP} algorithm that does not take noise into account. In Figure \ref{fig:1bitalgorithm}, we summarize the NMSE performance of the one-bit algorithms. Figures 2(a) and 2(b) show the NMSE performance of one-bit CS algorithms as a function of {$1-\gamma$}, and $M$, respectively. From Figures 2(a) and 2(b), we see that the proposed algorithm has superior performance compared to \texttt{R1bcs}, \texttt{SignGamp}, \texttt{BIHT}, and \texttt{AdtGamp}. \texttt{BIHT} and \texttt{AdtGAMP} perform the worst. 
	Further, the \texttt{BIHT} algorithm does not account for the noise, which leads to poor performance. We note that the proposed algorithm performs better than the \texttt{R1bcs} algorithm, which is a Bayesian algorithm that is robust to sign-flip noise. Moreover, the \texttt{R1bcs} algorithm requires matrix inversion in the algorithm and is computationally expensive than the proposed algorithm. 
	From the first experiment, we conclude that accounting for both pre-quantization and post-quantization noise leads to improved reconstruction performance. In the following experiments, we consider the performance of \texttt{Noisy1bG} as the baseline and compare the performance of the SI based algorithms.

	\begin{figure*}[htb] 
		\centering
		\begin{subfigure}[b]{0.45\textwidth}
			\includegraphics[width = 1.\textwidth]{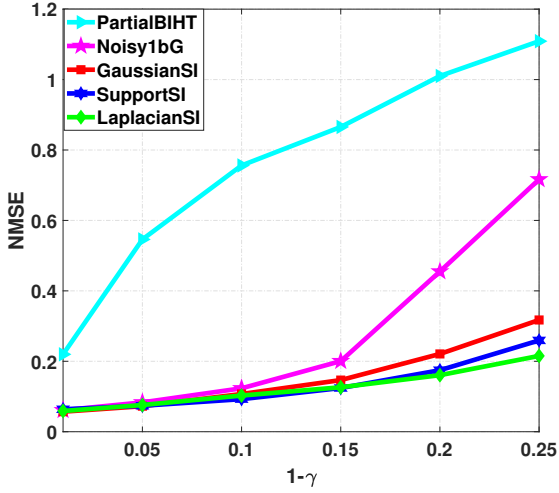}
			\caption{NMSE against bit-flip probability}
			\label{fig:gammaSI}
		\end{subfigure} 
		~ \begin{subfigure}[b]{0.45\textwidth}
			\includegraphics[width=1.\textwidth]{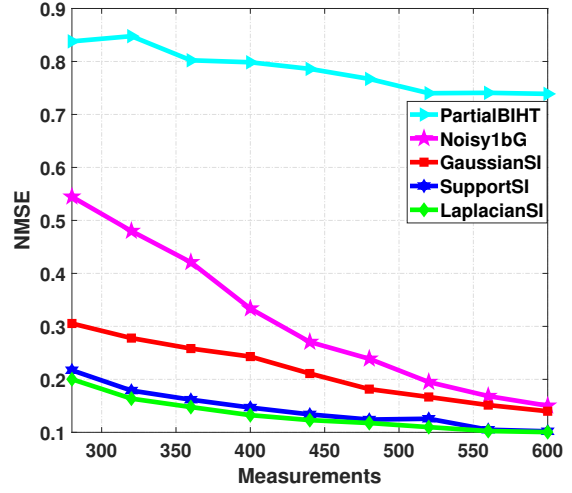}
			\caption{ \footnotesize NMSE as a function of $M$}
			\label{fig:MSI}
		\end{subfigure} 
		\caption{Comparison of reconstruction performance of the proposed methods in presence of SI when $N=200$, $\lambda = 0.15$.} \label{fig:1bitalgorithmSI}
	\end{figure*}

	In the second experiment, we study the reconstruction performance of sparse signals from their noisy one-bit compressed measurements when the receiver has access to some SI. We assume that the SI is erroneous. A small fraction of the elements in the support set of SI do not lie in the support set of the compressed signal. Further, we assume some additive noise present in the SI. The additive noise and the change of support are modeled by the Laplacian noise and the Gaussian noise in the proposed algorithms \texttt{LaplacianSI}, and \texttt{GaussianSI}, respectively. With the noisy SI at the receiver, Figure 3 demonstrates the reconstruction performance of the proposed algorithms. Figure 3(a) shows the performance of the proposed algorithms against sign-flip probability, and Figure 3(b) shows the performance of the proposed algorithms as a function of $M$. From both of these results, we conclude that all the proposed algorithms with SI perform better than the case when we do not have side-information. We emphasize that the \texttt{LaplacianSI} algorithm outperforms the GaussianSI algorithm. The error in support with the amplitude information between SI and the compressed signal can be modeled better by the Laplacian distribution than the Gaussian distribution. We further emphasize that the \texttt{SupportSI} algorithm only considers the support information as the side-information. We see that \texttt{SupportSI} performs better than the \texttt{GaussianSI} algorithm. As the change in support is difficult to model by Gaussian noise, we claim that the poor performance of \texttt{GaussianSI} is due to the modeling error.  
	\begin{figure}[htb]
		\centering
		\includegraphics[width = 0.4750\textwidth]{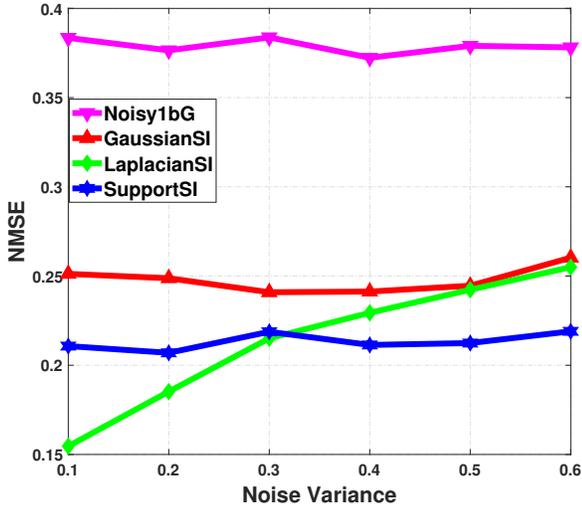}
		\caption{ Comparison of the effect of noise in SI on the reconstruction performance of proposed algorithms when  $\lambda$ = 0.1, $1-\gamma = 0.15,$ and $v = 0.15,$}
		\label{fig:fixedSI}
	\end{figure}  
	
	Third, we consider the effect of noise in SI on the reconstruction performance from one-bit measurements. Like in the second experiment, the SI at the receiver has a fraction of elements in its support set, which are not in the support of the compressed signals. Further, the amplitudes of the SI are corrupted by additive noise. In the experiment,  10\% of the elements in the support set of SI are not in the support set of the compressed signal. Further, we use Gaussian noise as the additive noise in the SI. In Figure 4, we plot the results of the experiment. It is evident that the performance of algorithms \texttt{Noisy1bG} and \texttt{SupportSI} is relatively constant for different values of the variance of additive noise. For the \texttt{SupportSI} algorithm, we assume that the knowledge of support does not change with the additive noise; hence it does not affect the performance of the algorithm. Since the Gaussian density could not model the sparse nature of the noise vector well, the performance of the \texttt{GaussianSI} algorithm  is worse than \texttt{SupportSI} algorithm for all values of the noise variance. The performance of the \texttt{LaplacianSI} degrades with the increase in the noise in SI. Note that, the performance of \texttt{LaplacianSI} is worse than \texttt{SupportSI} when the noise in the SI is above a certain level. Hence, using support information, if available, is better than using the entire SI signal when the signal to noise ratio of SI is small.

	In the final experiment, we consider the case where the support of the observed sparse signal changes slowly over time. 
	In the simulation, we generate a sequence of sparse signals such that 10\% of the support changes between two consecutive time instants. For the first time instant, the non-zero elements are generated from an i.i.d. Gaussian distribution with mean zero and variance 5.5. We then obtain the amplitudes of the indices that continue to be in the support set of the signal by adding a random vector with zero mean and a small variance generated from an i.i.d. Gaussian distribution. For the indices that are not in the support set of the signal at the previous time instant, the amplitudes are generated from an i.i.d. Gaussian distribution with mean zero and variance 5.5. The receiver has access to noisy one-bit measurements of these signals. The receiver estimates the sparse signal at the first time instant using the \texttt{Noisy1bG} algorithm. This estimate of the sparse signal is now fed to the \texttt{GaussianSI} and the \texttt{LaplacianSI} algorithms as the SI. Using this SI, the proposed algorithms estimate the compressed signal. In the next iteration, \texttt{GaussianSI} and \texttt{LaplacianSI} use their estimates of the previous time instant signal as the SI and estimate the compressed signal. Figure 5 shows the NMSE performance of the proposed algorithms. We can see that \texttt{LaplacianSI} performs better than the \texttt{GaussianSI} and \texttt{Noisy1bG} algorithms. 
	The \texttt{GaussianSI} algorithm, though worse than the \texttt{LaplacianSI} algorithm, performs better than the \texttt{Noisy1bG} algorithm. Hence we conclude that when the support of the signal changes slowly over time, using the signal reconstructed at the previous time instant as SI leads to improved performance than just using a one-bit reconstruction algorithm.      
	
	\begin{figure}
		\centering
		\includegraphics[width=0.9\linewidth]{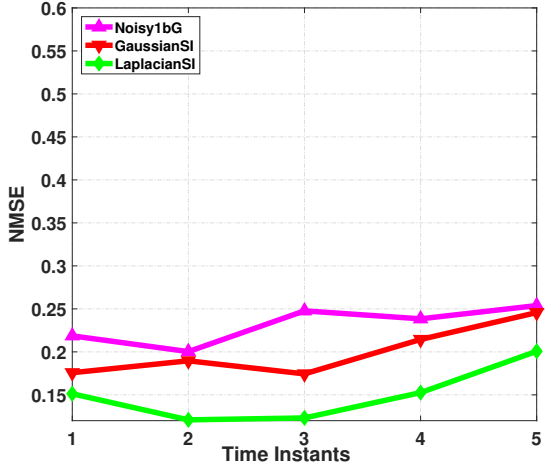}
		\caption{Comparison of reconstruction performance of the proposed algorithms when $\lambda$ = 0.1, $1-\gamma = 0.15,$ and $v = 0.15, N = 200,$ and $M= 600$}
		\label{fig:resdynsi}
	\end{figure}
	\section{Conclusion}
	In this work, we developed signal reconstruction algorithms from one-bit measurements using the generalized approximate message passing (\texttt{GAMP}) framework considering a generalized noisy measurement process. We then considered the scenario when side-information is available at the receiver. We developed two different algorithms to take into account SI which has either support information only or support and amplitude information. We derived closed-form expressions for \texttt{GAMP} estimation functions for all the proposed algorithms. We showed that by incorporating SI, we can improve the reconstruction performance in terms of NMSE. Further, we showed that the difference between the signal and the side information is better modeled by the Laplacian noise than Gaussian noise. We used the EM algorithm to estimate the noise parameter that governs our SI model adaptively from one-bit measurements and the side-information.
	Future work can consider extending the given algorithms to centralized and decentralized settings, especially when different nodes in a network have access to SI.
	\section*{Appendix A}
	\subsubsection*{Derivation of (\ref{PI1})}
	\begin{align*}
	PI_0(v,\widehat{p},\tau^p) & = \int \Phi(x/\sqrt{v}) \mathcal{N}(x;\widehat{p},\tau^p) dx \\
	& = \int_{-\infty}^{\infty} \Bigg[\int_{-\infty}^{x}  N(t|0,v) dt \Bigg] \mathcal{N}(x;\widehat{p},\tau^p) dx
	\end{align*}
	Using change of variable as $u = t-x + \widehat{p}$ and $w = x- \widehat{p},$ and changing the order of the integration, we get
	\begin{align*}
	&PI_0(v,\widehat{p},\tau^p) \\
	&=  \frac{1}{2\pi \sqrt{v \tau^p}}  \int_{-\infty}^{\tau^p} \int_{-\infty}^{\infty} \exp \Bigg\{- \Bigg[\frac{(u+w)^2}{2v} + \frac{u^2}{2 \tau^p}\Bigg] \Bigg\} dw~ du \\
	&=  \frac{1}{2\pi \sqrt{v \tau^p}} 
	\\& 
	\int_{-\infty}^{\tau^p} \int_{-\infty}^{\infty} \exp\Bigg\{ -\frac{1}{2}  \begin{bmatrix}
	w  \\
	u  
	\end{bmatrix}^T \begin{bmatrix}
	\frac{1}{2v} + \frac{1}{2 \tau^p} & \frac{1}{2v\tau^p}\\
	\frac{1}{2v\tau^p} & \frac{1}{2 \tau^p}
	\end{bmatrix} \begin{bmatrix}
	w \\
	u 
	\end{bmatrix}\Bigg\} dw du \\
	&=  \int_{-\infty}^{\tau^p} \int_{-\infty}^{\infty} \mathcal{N}\Bigg(\begin{bmatrix}
	w  \\
	u  
	\end{bmatrix} \Bigg{|} \mathbf{0}, \begin{bmatrix}
	\tau^p & -\tau^p\\
	-\tau^p & v + \tau^p
	\end{bmatrix} \Bigg) dw ~du \\
	&= \int_{- \infty}^{\mu} \mathcal{N}(u;0, v + \tau^p) du\\
	&= \Phi(\frac{\widehat{p}}{\sqrt{v + \tau^p}})
	\end{align*}
	The above expression represents marginalization of $w$ in the bi-variate normal density which is followed by the integration over $\left(\infty ~~ \tau^p\right].$ From the property of bivariate Gaussian distribution, marginalization of the bivariate normal density results in normal distribution. The mean and variance can be shown to be zero and $v+\tau$.  
	Next, consider the equality
	\begin{align}
	\int \Phi(x/\sqrt{v}) \mathcal{N}(x;\widehat{p},\tau^p) dx = \Phi\Big( \frac{\widehat{p}}{\sqrt{v + \tau^p}}\Big) \label{eqality}
	\end{align}
	Differentiating both side of (\ref{eqality}) with respect to $\widehat{p}$, we get,
	\begin{align}
	\begin{split}
	&\int \frac{x-\widehat{p}}{\tau^p}\Phi(x/\sqrt{v}) \mathcal{N}(x;\widehat{p},\tau^p) dx = \frac{\mathcal{N}(\frac{\widehat{p}}{\sqrt{v + \tau^p}})}{\sqrt{v + \tau^p}} \\
	&PI_1 = \widehat{p} \Phi\Big( \frac{\widehat{p}}{\sqrt{v + \tau^p}} \Big) +  \tau^p \frac{\mathcal{N}(\frac{\widehat{p}}{\sqrt{v + \tau^p}})}{\sqrt{v + \tau^p}}
	\end{split} \label{PI1}
	\end{align}
	Finally, differentiating both side of (\ref{PI1}) with respect to $\widehat{p}$, we get
	\begin{align*}
	&\int x\frac{x-\widehat{p}}{\tau^p} \Phi(x/\sqrt{v})\mathcal{N}(x;\widehat{p},\tau^p)) \\& = \Phi(\frac{\widehat{p}}{\sqrt{v+\tau^p}}) + \frac{\widehat{p}\mathcal{N}(\frac{\widehat{p}}{\sqrt{v+\tau^p}})}{\sqrt{v+\tau^p}} + \tau^p \widehat{p} \frac{\mathcal{N}(\frac{\widehat{p}}{\sqrt{v+\tau^p}})}{(v+\tau^p)^{1.5}} 
	\\&
	\implies PI_2 = \tau^p \Phi(\frac{\widehat{p}}{\sqrt{v+\tau^p}}) + \widehat{p} PI_1   +\frac{\tau^p ~\widehat{p}~ v \mathcal{N}(\frac{\widehat{p}}{\sqrt{v+\tau^p}})}{(v+\tau^p)^{1.5}} 
	\end{align*}
	
	\section*{Appendix B}
	\noindent
	Derivation of Results 1:
	
	\noindent Let $P_\mathcal{X}(x) \propto \mathcal{N}(x;m_G,v_G) \times  \mathcal{L}(x;m_L,v_L)$ be a probability density function. We compute mean of $x$ as
	\begin{flalign*}
	&\mathbb{E}[x] = \int x P_\mathcal{X}(x) dx = \frac{1}{Z}\int x \mathcal{N}(x;m_G,v_G) \times  \mathcal{L}(x;m_L,v_L) \\
	& = \frac{1}{Z}\Bigg[\int_{-\infty}^{m_L} x \mathcal{N}(x;m_G,v_G)\mathcal{L}(x;m_L,v_L)  ~dx \\&  \quad \quad \quad \quad \quad \quad  + \int_{m_L}^{\infty} x \mathcal{N}(x;m_G,v_G)\mathcal{L}(x;m_L,v_L) ~dx \Bigg]
	\end{flalign*} 
	After some algebraic steps,
	\begin{flalign*}
	&\frac{1}{Z}\Bigg[\int_{-\infty}^{m_L} x \mathcal{N}(x;m_G,v_G)\mathcal{L}(x;m_L,v_L)\bigg] 
	\\& = \frac{C_1}{Z} \int_{-\infty}^{m_L} x \mathcal{N}(x;m_G+\frac{v_G}{2v_L}, v_G) ~dx
	\\&= \frac{C_1}{Z} I_1(m_L; m_G+\frac{v_G}{2v_L}, v_G),
	\end{flalign*}
	where $C_1 = \frac{1}{4 v_L} \exp(-\frac{1}{2v_L}({m_L-m_G-\frac{v_G}{4v_L}})$. Similarly,
	\begin{align*}
	&\frac{1}{Z}[\int_{-m_L}^{\infty} x  \mathcal{N}(x;m_G,v_G)\mathcal{L}(x;m_L,v_L) \\ 
	& = \frac{C_2}{Z} \int_{-m_L}^{\infty} x \mathcal{N}(x;m_G-\frac{v_G}{2v_L}, v_G)~dx \\
	&= \frac{C_2}{Z} \Big(m_G-\frac{v_G}{2v_L} - I_1\Big(m_L; m_G-\frac{v_G}{2v_L}, v_G\Big) \Big),
	\end{align*} 
	where $C_2 = \frac{1}{4 v_L} \exp(-\frac{1}{2v_L}\big(-m_L+m_G-\frac{v_G}{4v_L}\big))$. The normalization constant, $Z$, can be evaluated as
	\begin{align}\label{normConst}
	\begin{split}
	& Z = \int \mathcal{N}(x;m_G,v_G)\mathcal{L}(x;m_L,v_L)dx \\
	&= C_1 \int_{-\infty}^{m_L} \mathcal{N}(x;m_G +\frac{v_G}{2v_L}, v_G) +  C_2 \int_{m_L}^{\infty} \mathcal{N}(x;m_G-\frac{v_G}{2v_L}, v_G) \\
	&= C_1 \Phi(\frac{m_L- (m_G+ \frac{v_G}{2v_L})}{\sqrt{v_G}}) + C_2 \Big(\Phi\Big(-\frac{m_L- \Big(m_G- \frac{v_G}{2v_L}\Big)}{\sqrt{v_G}} \Big) \Big)
	\end{split}
	\end{align}
	Next, we compute $\mathbb{E}[x^2]$ using the definition
	\begin{align}
	\begin{split}
	&\mathbb{E}[x^2] = \int  x^2 \mathcal{N}(x;m_G,v_G)\mathcal{L}(x;m_L,v_L)dx \\
	&= C_1 \int_{-\infty}^{m_L}  x^2 \mathcal{N}(x;m_G+\frac{v_G}{2v_L}, v_G) 
	\\& 
	\quad \quad \quad \quad \quad \quad +   C_2 \int_{-m_L}^{\infty}  x^2 \mathcal{N}(x;m_G-\frac{v_G}{2v_L}, v_G) \\&= \frac{C_1}{Z} I_2(m_L; m_G+ \frac{v_G}{2v_L}, v_G) \\&+\frac{C_2}{Z} \Big(v_G + \Big(m_G-\frac{v_G}{2v_L})^2 - I_2(m_L; m_G-\frac{v_G}{2v_L}, v_G \Big)\Big).
	\end{split}
	\end{align}
	Using these results, we derive $\mathbb{E}_{\mathcal{X}_n|\bm{\mathcal{Y}},\widetilde{X}}[x_n|\mathbf y, \widetilde{x}_n;\widehat{r}_n,\tau^r_{n}] $, and  $\mathbb{E}_{\mathcal{X}_n|\bm{\mathcal{Y}};\widetilde{X}}[x^2_n|\mathbf y, \widetilde{x}_n,\widehat{r}_n,\tau^r_{n}].$ Note that, the receiver has access to side-information which is assumed to be the actual signal corrupted by Laplacian noise.
	\begin{equation*}
	\widehat{\mathcal R}_n = \mathcal{X}_n + \mathcal{V}_n, ~~~~~~~~~    \widetilde{\mathcal X}_n = \mathcal{X}_n + \mathcal{W}_n 
	\end{equation*}
	where $\mathcal{V}_n \sim \mathcal{N}(0, \tau^r_{n})$,  $\mathcal{W}_n \sim \mathcal{N}(0,v_s)$ and $\mathcal{X}_n \sim p_{\mathcal{X}_n}(x_n)$ are independent.
	The GAMP algorithm approximates the marginal posterior distribution as 
	\begin{align*}
	&p_{\mathcal{X}_n|\bm{\mathcal{Y}}}(x_n|\mathbf{y};\widehat{r}_n,\tau^r_{n}) 
	\\&  \quad \quad \quad \quad \quad \quad \quad =\frac{p_{\mathcal{X}_n}(x_n)\mathcal{N}(x_n;\widehat{r}_n,\tau^r_{n}) \mathcal{L}(x_n;\widetilde{x}_n, 2v_s)}{\int_{x_n}p_{\mathcal{X}_n}(x_n)\mathcal{N}(x;\widehat{r}_n,\tau^r_{n}) \mathcal{L}(x_n;\widetilde{x}_n, 2v_s)}.
	\end{align*}

	\noindent \textit{Normalization Constant:}
	\begin{align*}
	\begin{split}
	&Z_n^l = \int_{-\infty}^{\widetilde{x}_n} \mathcal{N}(x_n;\widehat{r}_n, \tau_n^r)  ~ p_{\mathcal{X}_n}(x_n) \frac{1}{4v_s}\exp(\frac{-|x_n- \widetilde{x}_n|}{2v_s}) dx \\
	&= \int_{-\infty}^{\infty}   \frac{1-\lambda}{4v_s}\mathcal{N}(x_n;\widehat{r}_n, \tau_n^r)  ~\exp(\frac{ |x_n- \widetilde{x}_n|}{2v_s}) \delta(x_n) dx_n  
	\\& +\int_{- \infty}^{\infty}  \frac{\lambda}{4v_s} \mathcal{N}(x_n;\widehat{r}_n, \tau_n^r) \mathcal{N}(x_n;0, v_x) ~\exp(-\frac{|x_n- \widetilde{x}_n|}{2v_s}) dx_n
	\end{split}
	\end{align*}
	Using Gaussian product rule, $\mathcal{N}(x_n;\widehat{r}_n, \tau_n^r) \mathcal{N}(x_n;0, v_x) =  \mathcal{N}(0;\widehat{r}_n, \tau_n^r+v_x) \mathcal{N}(x;\frac{v_x \widehat{r}_n}{v_x+\tau_n^r}, \frac{v_x\tau_n^r}{v_x+\tau_n^r})$, and (\ref{normConst}), we get
	\begin{align*}
	& = \frac{1- \lambda}{4v_s\sqrt{2\pi \tau_n^r}} \exp(-\frac{\widehat{r}_n^2}{2\tau_n^r} - \frac{|\widetilde{x}_n|}{2 v_s}) 
	+ \lambda \phi\big(\frac{\widehat{r}_n}{\sqrt{\tau_n^r+v_x}}\big)
	\\& \Big(C_{1,n} \Phi(\frac{m^l_n- (m^g_n+ \frac{v^g_n}{2v^l_n})}{\sqrt{v^g_n}}) + C_{2,n} \Bigg(1-\Phi(\frac{m^l_n- (m^g_n- \frac{v^g_n}{2v^l_n})}{\sqrt{v^g_n}}) \Bigg)
	\end{align*}
	where $m^g_n = \frac{v_x \widehat{r}}{v_x+\tau^r}$, $v^g_n  = \frac{v_x \tau^r}{v_x+\tau^r}, m^l_n = \widetilde{x}_n, v^l_n = v_s$. 
	$C_{1,n}$ and $C_{2,n}$ depend on parameters $m^g_n , v^g_n , m^l_n ,$ and $v^l_n. \newline $

	\noindent\textit{Derivation of $\mathbb{E}_{\mathcal{X}_n|\bm{\mathcal{Y}},\widetilde{X}_n}[x_n|\mathbf y, \widetilde{x}_n;\widehat{r}_n,\tau^r_{n}]: $}
	\begin{align*}
	\begin{split}
	\quad&\mathbb{E}_{\mathcal{X}_n|\bm{\mathcal{Y}},\widetilde{X}}[x_n|\mathbf y, \widetilde{x}_n;\widehat{r}_n,\tau^r_{n}] 
	\\ \quad \quad& = \frac{1}{Z_n^l} \int x_n \mathcal{N}(x_n;\widehat{r}_n, \tau_n^r) \mathcal{N}(x_n;0,v_x) \mathcal{L}(x_n;\widetilde{x}_n,2v_s) dx_n\\
	& = \frac{\lambda \phi\big(\frac{\widehat{r}_n}{\sqrt{\tau_n^r+v_x}}\big)}{Z_n^l}  \Bigg(C_{1,n} I_1(m^l_n; m^g_n +\frac{v^g_n }{2v^l_n }, v^g_n ) + 
	\\& 
	\quad \quad \quad \quad \quad 
	C_{2,n} \Big(m^g_n -\frac{v^g_n }{2v^l_n } - I_1(m^l_n ; m^g_n-\frac{v^g_n}{2v^l_n}, v^g_n ) \Big) \Bigg)
	\end{split}
	\end{align*}
	
	\noindent\textit{Derivation of $\mathbb{E}_{\mathcal{X}_n|\bm{\mathcal{Y}},\widetilde{X}}[x^2_n|\mathbf y, \widetilde{x}_n,\widehat{r}_n,\tau^r_{n}]:$}
	\begin{align*}	\label{}
	\begin{split}
	&\mathbb{E}_{\mathcal{X}_n|\bm{\mathcal{Y}},\widetilde{X}}[x^2_n|\mathbf y, \widetilde{x}_n;\widehat{r}_n,\tau^r_{n}]\\ &= \frac{\lambda}{Z_n^l}\int x_n^2 \mathcal{N}(x|\widehat{r}, \tau_n^r) \mathcal{N}(x;0,v_x) \mathcal{L}(x_n;\widetilde{x}_n,2v_s) dx_n\\
	& = \frac{\lambda \phi\big(\frac{\widehat{r}_n}{\sqrt{\tau_n^r+v_x}}\big)}{Z_n^l} \Bigg(C_{1,n} I_2\Big(m^l_n; m^g_n +\frac{v^g_n }{2v^l_n }, v^g_n \Big) + 
	\\& 
	C_{2,n} \Big( \Big(m^g_n -\frac{v^g_n }{2v^l_n }\Big)^2+ v^g_n - I_2\Big(m^l_n; m^g_n -\frac{v^g_n }{2v^l}, v^g_n \Big) \Bigg)\end{split}
	\end{align*}

	\section*{Appendix C}
	\subsection*{Derivation of $I_0(\widetilde{x}; m, \tau),  I_1(\widetilde{x}; m, \tau),$ and  $I_2(\widetilde{x}; m, \tau)$}
	We know that $I_0(\widetilde{x}; m, \tau) = \int_{-\infty}^{\widetilde{x}} \mathcal{N}(x;m,\tau) dx = \Phi (\frac{\widetilde{x}-m}{\sqrt{\tau}}).$ Differentiating $ I_0(-\infty, \widetilde{x}; m, \tau)$ with respect to $m$, we get
	
	\begin{align}
	\begin{split}
	& \frac{\partial I_0(\widetilde{x}; m, \tau) }{\partial m} =  \int_{-\infty}^{\widetilde{x}} \frac{x-m}{\tau}    
	\mathcal{N}(x;m,\tau) dx 
	\\ &
	= \frac{1}{\tau} \Big\{\int_{-\infty}^{\widetilde{x}} x \mathcal{N}(x;m,\tau) dx - m \Phi(\frac{\widetilde{x} - m}{\sqrt{\tau}}) \Big\} 
	\\ & 
	\Rightarrow  I_1(\widetilde{x}; m, \tau) = 
	m \Phi(\frac{\widetilde{x} - m}{\sqrt{\tau}}) - \sqrt{\tau }\phi(\frac{\widetilde{x}-m}{\sqrt{\tau}}).\end{split} \label{I1}
	\end{align}
	
	\noindent Differentiating $I_0(\widetilde{x} ; m, \tau)$ twice with respect to $m$, and following steps similar to those in (\ref{I1}), we get
	\begin{align}
	\begin{split}
	&I_2(\widetilde{x}; m, \tau)  = m I_1(\widetilde{x}; m, \tau) + \tau~ \Phi(\frac{\widetilde{x}-m}{\sqrt{\tau}}) 
	\\& ~~~~~~~~~~~~~~~~~~~~~~~~ - \widetilde{x} \sqrt{\tau} \phi(\frac{\widetilde{x}-m}{\sqrt{\tau}}). \end{split} 
	\end{align}

	\section*{Appendix D}
	\noindent \textit{Derivation of Result 3:}
	
	\noindent The receiver has access to side-information which is assumed to be the actual signal corrupted by Gaussian noise.
	\begin{equation*}
	\widehat{\mathcal R}_n = \mathcal{X}_n + \mathcal{V}_n, ~~~~~~~~~    \widetilde{\mathcal X}_n = \mathcal{X}_n + \mathcal{W}_n 
	\end{equation*}
	where $\mathcal{V}_n \sim \mathcal{N}(0, \tau^r_{n})$,  $\mathcal{W}_n \sim \mathcal{N}(0,v_s)$ and $\mathcal{X}_n \sim p_{\mathcal{X}_n}(x_n)$ are independent.\newline
	The GAMP algorithm approximates the marginal posterior distribution as 
	\begin{align*}
	&p_{\mathcal{X}_n|\bm{\mathcal{Y}}}(x_n|\mathbf{y};\widehat{r}_n,\tau^r_{n}) 
	\\&  \quad \quad \quad \quad \quad \quad \quad =\frac{p_{\mathcal{X}_n}(x_n)\mathcal{N}(x_n;\widehat{r}_n,\tau^r_{n}) \mathcal{N}(x_n;\widetilde{x}_n, v_s)}{\int_{x_n}p_{\mathcal{X}_n}(x_n)\mathcal{N}(x;\widehat{r}_n,\tau^r_{n}) \mathcal{N}(x_n;\widetilde{x}_n, v_s)}.
	\end{align*}
	Next, we derive the posterior mean and variance. \\ \newline 
	\noindent\textit{Derivation of $\mathbb{E}_{\mathcal{X}_n|\bm{\mathcal{Y}},\widetilde{X}_n}[x_n|\mathbf y, \widetilde{x}_n;\widehat{r}_n,\tau^r_{n}] $}
	\begin{align} \label{eqn:GauMean}
	\begin{split}
	&\mathbb{E}_{\mathcal{X}_n|\bm{\mathcal{Y}},\widetilde{X}}[x_n|\mathbf y, \widetilde{x}_n;\widehat{r}_n,\tau^r_{n}] 
	\\& = \frac{1}{Z_n^g} \int x_n \mathcal{N}(x_n|\widehat{r}_n, \tau_n^r) \mathcal{N}(x_n|0,v_x) \mathcal{N}(x_n;\widetilde{x}_n,v_s) dx_n\\
	& = \frac{\lambda \phi\big(\frac{\widehat{r}_n}{\sqrt{\tau_n^r+v_x}}\big)}{Z_n^g}   \int x_n \mathcal{N}(x_n;\frac{\widehat{r}_n v_x}{v_x+\widehat{r}_n} , \frac{v_x \tau^r_n}{v_x+\tau^r_n}) \mathcal{ N}(x_n;\widetilde{x}_n , v_s) dx_n
	\\& = \frac{\lambda \phi\big(\frac{\widehat{r}_n}{\sqrt{\tau_n^r+v_x}}\big)}{Z_n^g}  \mathcal{N}(x_n;\frac{\widehat{r}_n v_x}{v_x+\widehat{r}_n} , \frac{v_x \tau^r_n}{v_x+\tau^r_n} + v_s) 
	\\& \quad \quad  \quad \int x \mathcal{N}(x_n;\frac{\widehat{r}_n v_x v_s+ v_x \tau_n^r \widetilde{x}_n  }{v_x\tau_n^r + v_x v_s + v_s \tau^r_n}, \frac{v_s v_x\tau_n^r}{v_x\tau_n^r + v_x v_s + v_s \tau^r_n}) dx_n
	\\&  = \frac{\lambda \phi\big(\frac{\widehat{r}_n}{\sqrt{\tau_n^r+v_x}}\big)}{Z_n^g} \mathcal{N}(0;\frac{\widehat{r}_n v_x}{v_x+\widehat{r}_n} , \frac{v_x \tau^r_n}{v_x+\tau^r_n} + v_s)   \frac{\widehat{r}_n v_x v_s+ v_x \tau_n^r \widetilde{x}_n }{v_x\tau_n^r + v_x v_s + v_s \tau^r_n},
	\end{split}
	\end{align}
	where $Z^g_n$ is the normalization constant. The normalization constant is evaluated as 
	\begin{align} \label{eqn:GauCnst}
	\begin{split}
	&Z_n^g = \int ~ p_{\mathcal{X}_n}(x_n) ~ \mathcal{N}(x_n;\widehat{r}_n, \tau_n^r) ~ \mathcal{N}(x;\widetilde{x}_n, v_x) dx 
	\\& = \int (1-\lambda)\mathcal{N}(x_n;\widetilde{x}_n, \tau_n^r)  \mathcal{N}(x_n;\widehat{r}_n, \tau_n^r) \delta(x)+ 
	\\& \quad \quad \quad \lambda \mathcal{N}(x_n;\widehat{r}_n, \tau_n^r) ~ \mathcal{N}(x;\widetilde{x}_n, v_x) \mathcal{N}(x_n;0,v_x) dx
	\\& = (1-\lambda)\mathcal{N}(0;\widetilde{x}_n, \tau_n^r)  \mathcal{N}(0;\widehat{r}_n, \tau_n^r) 
	\\&  \quad \quad \quad + \lambda \mathcal{N}(0;\frac{\widehat{r}_n v_x}{v_x+\widehat{r}_n} , \frac{v_x \tau^r_n}{v_x+\tau^r_n} + v_s) \phi\big(\frac{\widehat{r}_n}{\sqrt{\tau_n^r+v_x}}\big)  
	\end{split}
	\end{align}
	Replacing  (\ref{eqn:GauCnst}) in (\ref{eqn:GauMean}), and with some algebraic steps, we can show that
	\begin{align*}
	\mathbb{E}_{\mathcal{X}_n|\bm{\mathcal{Y}},\widetilde{X}}[x_n|\mathbf y, \widetilde{x}_n;\widehat{r}_n,\tau^r_{n}] = \pi^g_{n} \frac{\widehat{r}_n v_s v_x + v_x \tau^r_{n} \widetilde{x}}{v_x\tau^r_{n} + \tau^r_{n}v_s + v_s v_x} \triangleq \pi^g_{n} m^g_n
	\end{align*}
	where $\pi^g_n = \frac{ \lambda }{\lambda + (1-\lambda) Z_n} $ and
	$Z_n = \frac{\mathcal{N}(0;\widehat{x}, \tau^r_{n}) \mathcal{N}(0;\widehat{x}, \tau^r_{n})}{\mathcal{N}(0;\widehat{r}_n, v_x + \tau^r_{n}) \mathcal{N}(0;\frac{\widehat{r}_n v_x}{v_x+\tau^r_{n}}-\widetilde{x}, \frac{\tau^r_{n}v_x}{v_x + \tau^r_{n}} + v_s)}. \newline$
	
	\noindent\textit{Derivation of $\mathbb{E}_{\mathcal{X}_n|\bm{\mathcal{Y}},\widetilde{X}}[x^2_n|\mathbf y, \widetilde{x}_n,\widehat{r}_n,\tau^r_{n}]$\newline}
	
	\noindent Using the definition of second-order moment and following the similar algebraic steps we can write 
	\begin{align}
	\begin{split}
	&\mathbb{E}_{\mathcal{X}_n|\bm{\mathcal{Y}},\widetilde{X}}[x^2_n|\mathbf y, \widetilde{x}_n,\widehat{r}_n,\tau^r_{n}] 
	\\& = \frac{1}{Z_n^g} \int x^2_n~\mathcal{N}(x_n;\widehat{r}_n, \tau_n^r) \mathcal{N}(x_n;0,v_x) \mathcal{N}(x_n;\widetilde{x}_n,v_s) dx_n
	\\& = \pi^g_{n} \Big( \frac{v_s v_x \tau^r_{n}}{v_x\tau^r_{n} + \tau^r_{n}v_s + v_s v_x} + (m^g_n)^2\Big)
	\end{split}
	\end{align} 
	
	Using the posterior first-order and second-order moments, the posterior variance can be expressed as 
	\begin{align*}
	&\text{var}_{\mathcal{X}_n|\bm{\mathcal{Y}},\widetilde{X}_n}[x_n|\mathbf y, \widetilde{x}_n;\widehat{r}_n,\tau^r_{n}]
	\\& \quad \quad \quad \quad = \pi^g_{n} \frac{v_s v_x \tau^r_{n}}{v_x\tau^r_{n} + \tau^r_{n}v_s + v_s v_x} + \pi^g_{n} (1-\pi^g_{n}) (m^g_n)^2
	\end{align*}
	
	\bibliographystyle{IEEEtran}
	\bibliography{IEEEabrv,bib2}
\end{document}